%% file: main.tex
\documentclass[preprint, 12pt]{elsarticle}

\usepackage{color}
\usepackage{multirow}
\usepackage{array}
\usepackage{amsmath,amsthm}
\usepackage{mathtools}
\usepackage{subfig}
\usepackage[export]{adjustbox}
\usepackage{epstopdf}
\usepackage[utf8]{inputenc}
\usepackage{balance}
\usepackage{amsmath,amssymb,amsfonts}
\usepackage{algorithmic}
\usepackage{graphicx}
\usepackage{textcomp}
\usepackage{xcolor}
\usepackage{array}
\usepackage{url}
\usepackage[english]{babel}
\usepackage{standalone}
\usepackage{enumitem}
\newtheorem*{remark}{Remark}

\usepackage{listings}

\usepackage{tikz}
\usetikzlibrary{positioning}
\usepackage{pgfplots}
\usepackage{pgfplotstable}

\journal{Simulation Modelling Practice and Theory}

\begin{document}

\begin{frontmatter}



\title{Fog to cloud and network coded based architecture: Minimizing data download time for smart mobility}


\author[1]{Goiuri Peralta\corref{cor1}}\ead{gperalta@ikerlan.es}
\author[1]{Pablo Garrido}
\author[1]{Josu Bilbao}
\author[2]{Ram\'{o}n Ag\"{u}ero}
\author[3]{Pedro M. Crespo}

\address[1]{Information and Communication Technologies Area, Ikerlan Technology Research Centre, Arrasate-Mondragón, Spain}

\address[2]{Dpt. of Communications Engineering, University of Cantabria, Santander, Spain}

\address[3]{Electronics and Communications Department, University of Navarra (Tecnun), Donostia-San Sebastián, Spain}

\cortext[cor1]{Corresponding author.}

\begin{abstract}
Industry 4.0 applications foster new business opportunities, but they also pose new and challenging requirements, such as low latency communications and highly reliable systems. They would likely exploit novel wireless technologies (5G), but it would also become crucial using architectures that appropriately support them. In this sense, the combination of fog and cloud computing represents a potential solution, since it can dynamically allocate the workload depending on the specific needs of each application. In this paper, our main goal is to provide a highly reliable and dynamic architecture, which minimizes the time that an end node or user, spends in downloading the required data. In order to achieve this, we have developed an optimal distribution algorithm that decides the amount of information that should be stored at, or retrieved from, each node, to minimize the overall data download time. Our scheme is based on various parameters and it exploits Network Coding (NC) as a tool for data distribution, as a key enabler of the proposed solution. We compare the performance of the proposed scheme with other alternative solutions, and the results show that there is a clear gain in terms of the download time.
\end{abstract}


\end{frontmatter}


\input{Introduction.tex}

\input{Related_work.tex}

\input{Proposal.tex}

\input{Results.tex}

\input{Conclusion.tex}

\section*{Acknowledgements}
This work has been partially supported by the Basque Government through the ADDISEND Elkartek program (Grant agreement no. KK-2018/00115), the DIGITAL Elkartek program (Grant agreement no. KK-2019/00095), the H2020 research framework of the European Commission under the ELASTIC project (Grant agreement no. 825473), and the Spanish Ministry of Economy and Competitiveness through the CARMEN project (TEC2016-75067-C4-3-R), the ADVICE project (TEC2015-71329-C2-1-R), the FIERCE project (RTI2018-093475-A-100) and the COMONSENS network (TEC2015-69648-REDC).

\input{Appendix.tex}

\bibliographystyle{elsarticle-num} 
\bibliography{bibliography}

\end{document}

%% file: Introduction.tex
\section{Introduction}
\label{intro}
The exponential growth of applications and services for the so-called Industry 4.0 paradigm, and the advocate of Industrial IoT (IIoT), are behind the appearance of new challenges \cite{Aijaz19,Stefanovic18, Boyes18}. Applications where real-time monitoring and processing are critical features for immediate decision making processes, such as robotic guidance, e-health, or smart mobility, pose stringent requirements. These are, among others, low latency communication, and high reliability and availability \cite{Gangakhedkar18,Thuemmler17}. 

\begin{figure}[!h]
\centering
\includegraphics[width=0.75\linewidth]{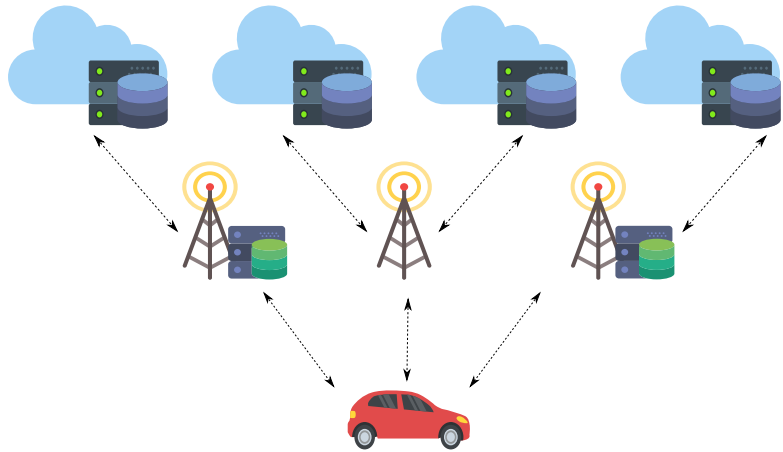}
\caption{Reference fog to cloud architecture.}
\label{fig:arquitectura}
\end{figure}

5G aims at coping with the requirements of Ultra-Reliable Low-Latency Communication (URLLC) systems, reducing the latency to 1~ms and providing up to 10~Gbps data rates \cite{Pocovi18,Cheng18,Li18}. Besides applying novel and advanced communication technologies, it is crucial to deploy an architecture that adequately supports these requirements. Fog computing \cite{Bonomi14,Chiang16} allows to perform early analytics and so reduce the latency for delay sensitive applications, while cloud computing \cite{Yue15,Georgakopoulos16} enables to store, process, and manage large amounts of data (Big Data). Thus, an architecture based on the combination of both fog and cloud computing paradigms (see Fig.~\ref{fig:arquitectura} for an illustrative example) can be an advantageous solution~\cite{Bittencourt18, Varshney17}, since it enables to provide the most suitable service by dynamically allocating the workload depending on the needs of each application.  

Taking smart mobility applications as our driving use-case, the main objective of this paper is to provide a highly reliable, yet flexible and dynamic architecture, which aims at minimizing the time that the end node or user, for instance a car, spends in downloading the required data. With such goal, we propose a combined fog/cloud architecture, as depicted in Fig.~\ref{fig:arquitectura}, where multiple nodes with different capabilities are available. For instance, fog nodes, due to their closeness to the end nodes, may provide lower communication latency, compared to the cloud environment. However, they might be characterized with a lower computational capacity, or larger costs. Under such circumstances, it might be more beneficial to use the cloud nodes. 

The first contribution of this paper is to model the system that represents the architecture shown in Fig.~\ref{fig:arquitectura}. We consider the following parameters: (1) the download bandwidth or rate, $Rb$; (2) the communication delay, $d_{link}$; and (3) the service time, i.e. the time required to process the data download request at a particular node, $d_{service}$. The values for $Rb$ and $d_{link}$ that have been used throughout the paper are based on the related literature \cite{Barcelo16, OpenSignal_1, OpenSignal_2}. In order to model $d_{service}$ we have applied a traditional $M/M/1$~\cite{kleinrock} queuing system. The second contribution of the paper is to develop an optimal distribution algorithm to minimize the data download time. It decides, based on the system parameters, the amount of information that should be stored in, or retrieved from, each node. We exploit Network Coding (NC) \cite{Ahlswede00, Fragouli06} as a mechanism for distributed data storage, which has been showed to bring relevant benefits~\cite{Dimakis10}. It allows reducing the amount of redundant data and so the download time \cite{Dimakis11,Fitzek14}, while ensuring the fault-tolerance of the system.

Based on the aforementioned model, we pose an optimization problem, which we solve to assess the performance of the proposed scheme in different scenarios, comparing it with that exhibited by other distribution methods. We have as well considered particularly difficult circumstances, characterized by poor connectivity or overload conditions, to assess whether the different schemes are able to adapt to such situations.

The remainder of this paper is organized as follows. In Section \ref{relwork}, we review the related work by introducing some of the benefits of the combination of cloud and fog computing for Industry 4.0 applications. We also discuss the potential benefits of exploiting NC techniques for the proposed distributed storage solution. After reviewing the existing works, we highlight the novel contributions of this paper. In Section \ref{system}, we describe the proposed scenario and our system model, depicting the parameters that have been introduced. Moreover, in Section \ref{min_problem}, we discuss the optimization problem that we pose to find the best data distribution strategy, with the objective of minimizing the total download time. In Section \ref{results}, we compare our proposal to other existing solutions, and analyze the impact of the system parameters. Finally, in Section \ref{conclu}, we conclude the paper, and we outline our future work.

%% file: Related_work.tex
\section{Related Work}
\label{relwork}

Cloud-based architectures are the most widely used in Industry 4.0 applications, due to their flexibility and efficiency,
 allowing both horizontal and vertical integration \cite{Yue15, Botta16}. However, it might not always meet the stringent requirements posed by Industry 4.0 or IIoT applications. Fog computing has been proposed as a means to overcome these limitations, and to extend cloud capabilities to the network edge~\cite{Bonomi14, Yi15, Baccarelli17}. It enables performing early analytics and closed-loop control with minimum latency. The combination between both computing paradigms (cloud and fog) can be thus highly beneficial. The advantages that the fog-cloud architecture can provide to smart city \cite{Sinaeepourfard17,Santos18}, e-health \cite{Farahani18}, and IoT \cite{Bittencourt18} scenarios have already been evinced in the available literature.

Authors in \cite{Sanchez18} propose a middleware that allows taking advantage of the capabilities of both paradigms, while overcoming typical problems of cloud-based systems, such as low latency, real-time, geo-distribution. Moreover, multi-cloud deployments have been also proposed as a solution to such issues \cite{Bessani13, Grozev14, Alshammari17}. The deployment of more than one cloud might indeed provide fault-tolerance against service outages, while system reliability is increased, since it becomes possible to store redundant data at the different clouds. Furthermore, application requirements can be better adapted to available resources and connectivity conditions.

Specifically, for smart mobility or connected cars, the advantages of combining fog and cloud computing paradigms have also been discussed in previous works~\cite{AutomationAlley,ThornTech}. Authors in \cite{Huang16} propose a distributed strategy for reliable real-time streaming in vehicular cloud-fog networks. In \cite{Pereira19}, the authors propose a generic architecture for the deployment of fog computing applications and services in a Vehicular Ad-Hoc Network (VANET) environment. Critical decisions that require almost real-time responses are better managed with fog computing, while cloud computing may be more suitable for advanced analytics, to bring insights that can be exploited in maintenance tasks, and reduce, for instance, the repair costs of a truck fleet.

The use of NC techniques has been shown to strengthen the benefits of these distributed systems. Due to its rateless nature, it ensures a fault-tolerant system with a more efficient redundancy. In addition, it allows the system to decouple itself from the underlying topology and scheduling approach \cite{Dimakis10}. Previous research, such as \cite{Saleh16}, develops a model to study different redundancy strategies, and to evaluate their performance on NC-based P2P video streaming systems. In \cite{Chen14}, authors show the potential of NC to ensure fault-tolerance in case of server failures in multi-cloud deployments. The use of NC techniques can be advantageous not only in a cloud environment, but also for fog deployments. Authors in \cite{Cabrera16} and \cite{Zhao18} study the distributed storage problem, particularly when data should be repaired with the remaining nodes in a fog deployment without newcomer nodes. They also design a practical testbed over Raspberry-Pi devices, using a NC implementation.

The main contribution of this paper is the proposal of a method that aims at minimizing the data download time for a fog-cloud architecture by optimally distributing the required data over the available system nodes. In previous works \cite{Sipos14, Sipos19}, the authors introduced a system that employs commercially available clouds to reliably store files. The original file is divided into a number of uncoded packets by the client, which are then linearly combined using RLNC to generate a larger number of coded packets. Such packets are then distributed by the client to $N$ clouds. The amount of information to store in (receive from) each node is then established by taken into account only the download rate of the storage node, i.e. finding whether a faster cloud should store a larger part of the data to improve download time. Opposed to this, we consider not only the download rate, but also the communication delay, and the service time of the available nodes. Then, we minimize the download time by optimally distributing the data based on such parameters.

%% file: Proposal.tex
\section{System model}
\label{system}

\begin{figure}[!h]
\centering
\vspace{-0.3cm}
\includegraphics[width=0.75\linewidth]{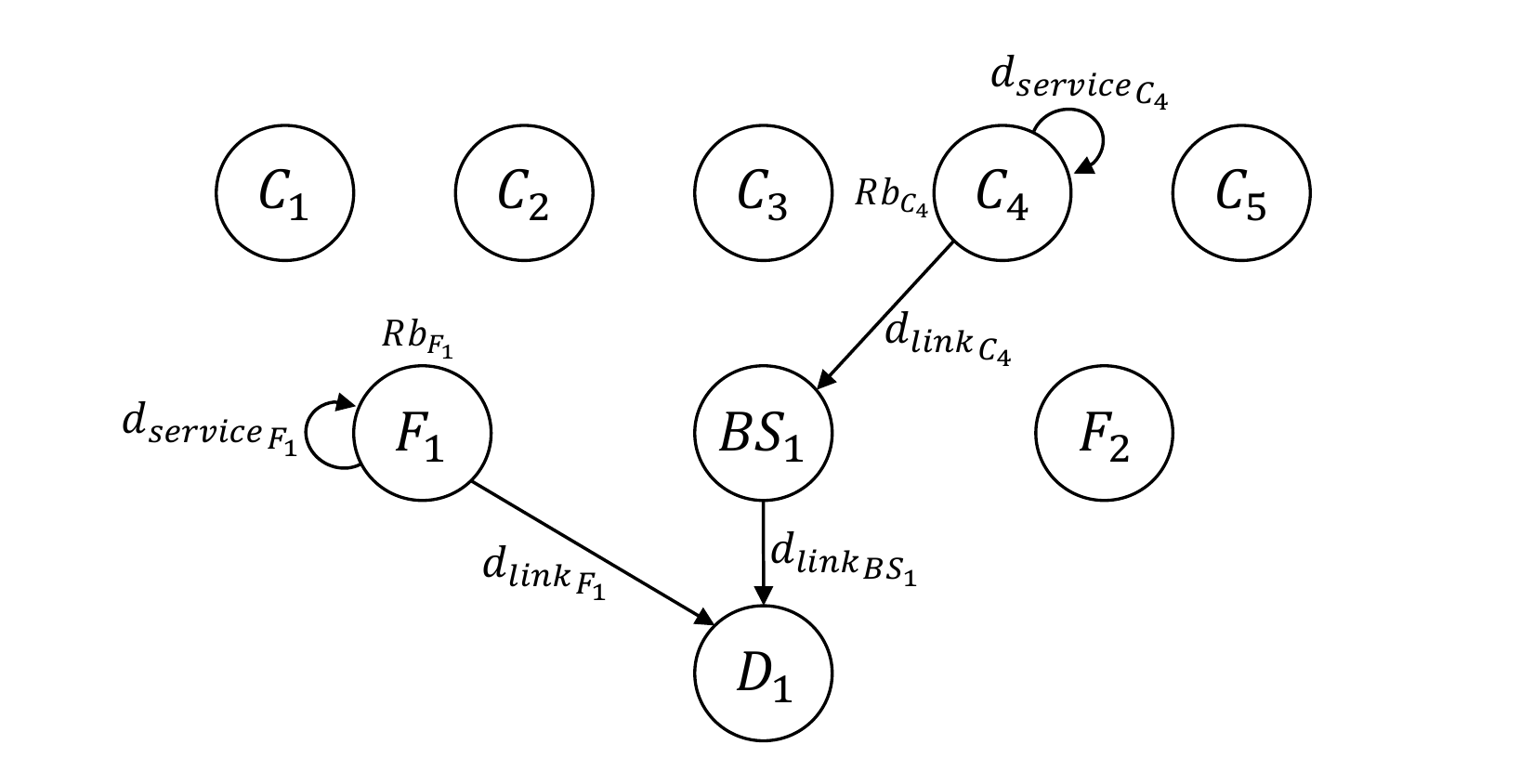}
\vspace{-0.3cm}
\caption{Illustrative example of the proposed system model, featuring an end device ($D_1$), five cloud ($C_x$) nodes, 2 fog ($F_x$) nodes, and one base station ($BS$).}

\label{fig:sistema}
\end{figure}

As was previously mentioned, the main objective of this paper is to provide a highly reliable and dynamic architecture based on the combination of both fog and cloud nodes, exploiting the benefits brought by NC for distributed storage. We have modeled such architecture and we have posed an optimization problem that seeks minimizing the overall download time. The system comprises a number of cloud servers and base stations (BS) (refer to Fig.~\ref{fig:arquitectura}). While some of the  BS are just used as an access element to communicate with the cloud environment, there are others that also include additional capabilities, such as data storage and processing. We will refer to them as fog nodes. This architectural approach is currently being considered in various scenarios, in particular~\cite{Pereira19} for vehicular applications. As depicted in Fig.~\ref{fig:sistema}, there is one end device, $D$, which downloads the required data from all the available clouds, $C$, and fog nodes, $F$. The access to a cloud node is done either through a legacy BS or a fog node\footnote{In this paper we assume that each of the cloud nodes is always connected to the best available access element, but this could be easily extended to consider situations where there might be some policies that avoid some clouds to use certain access elements. This is left for future work.}. We can actually exploit the information stored in all nodes by using NC, as will be discussed later.

We consider three parameters in our system model: the download rate, $Rb$, the communication delay, $d_{link}$, and the service time, i.e. the time required to process the data request, $d_{service}$. We use typical values of $Rb$ and $d_ {link}$ for 4G (Long Term Evolution, LTE) technology, based on bandwidth \cite{Barcelo16} and latency \cite{OpenSignal_1, OpenSignal_2} values taken from the literature. Regarding $d_{service}$, we have applied a $M/M/1$ queue~\cite{kleinrock} to model it, as  has been extensively used to model cloud servers \cite{Liu18, Wang18}. We thus assume that arrivals occur at a certain rate $\lambda$, following a Poisson process, while we model service times with an exponential distribution, having $\mu$ rate, with $t_s = \frac{1}{\mu}$ being the mean service time. Moreover, a single server serves customers one at a time, according to a FIFO policy and we assume the buffer to be large enough. Thus, there is no limit on the number of customers it can contain, and no losses are considered. The node occupation or load is calculated as $\rho = \frac{\lambda}{\mu}$, whose  value is always within the $[0, 1]$ interval. We assume that loads in the different nodes might be different, and would actually depend on their type, i.e. cloud or fog node. With this in mind, the time required to process a request can be defined as:

\begin{equation}
    d_{service}=\frac{t_s}{1-\rho}
\end{equation}

We focus on a snap-shot based analysis, where the system is studied for a particular situation. We nonetheless argue that it would be also possible to broaden the analysis, capturing the dynamics of the system, as we did in one of our previous work \cite{Peralta19}, where the availability of nodes changed depending on the cloud resource price. Since the goal of the paper was to assess the feasibility of fostering an optimum retrieval of the required data, we leave the analysis of the dynamic system for our future work.

As has been mentioned before, in our distributed storage system, we exploit NC, in particular Random Linear Network Coding (RLNC), which is possibly the most widespread NC solution, as a means to foster the distributed operation. 
RLNC breaks with the traditional store-and-forward transmission model, and it enables any intermediate node (router) to recombine incoming packets, which will be later decoded at the destination \cite{Ahlswede00, Ho06}. This is done by linearly combining the packets using randomly chosen coefficients from a finite field $\mathbb{F}_q$ of size $2^q$ ($q \gg 1$). The encoding operation of $M$ packets can be described as follows:
\begin{equation}
    p'_i=\sum_{j=1}^{M}c_{i,j} \cdot p_{j}
\end{equation}
where $[p_1,p_2,\ldots,p_M]$ are the original packets, $p'_i$ represents each coded packet, and $c_{j,i} \in \mathbb{F}_{q}$ are the coding coefficients. Thus, for each $p'_i$, the corresponding set of coding coefficients $[c_{i,1},c_{i,2},\ldots,c_{i,M}]$ actually conforms the coding vector. 

NC exploits the broadcast nature of the wireless medium, which facilitates node cooperation and it also provides significant benefits in terms of communication robustness, stability, throughput, and latency \cite{Bilbao16, Katti08, Xie15, Liu17}. Moreover, due to its rateless nature, it is not necessary to keep track of the coded packets that have been sent, and the receiver only needs to get enough linearly dependent packet combinations to recover the original data. This feature is fundamental for our proposed scheme, since it allows the destination to get packets regardless of the node used as a source. This greatly improves the storage efficiency in terms of data retrieval time and redundancy in distributed storage systems \cite{Chen14,Saleh16,Sipos14}. In a nutshell, the cloud and fog nodes send coded packets, which are built by randomly combining the original information. If we assume a large finite field, we can ensure, with high probability, that receiving any $k$ coded packets, the original information can be recovered.


By distributing the information among all nodes (cloud and fog), we can simultaneously exploit several information providers (clouds or fogs) to improve the performance~\cite{Fitzek14}. Furthermore, distributing the information yields a higher reliability, i.e. ensures fault-tolerance, of the system. As demonstrated by Dimakis \emph{et al.} in \cite{Dimakis10}, there is a trade-off curve that shows the relationship between the amount of information that is stored in each node $(\alpha)$ and the amount of information that needs to be transmitted among the nodes $(\gamma)$. These parameters may change depending on the goal of the application. For instance in our previous work \cite{Peralta19}, we analyzed the impact of different points of this trade-off curve to reduce the storage cost of a multi-cloud environment based on Amazon spot instances. 

As an example for the reliability requirement, we show below the minimum amount of information that must be stored in the system to ensure the reliability against the failure of one node, i.e. Minimum Storage Regenerating (MSR) configuration:

\begin{equation}
   \alpha_i \geq \frac{k}{N-1} 
\end{equation}
\noindent where $\alpha_i$ represents the percentage of information to store in each node and $N$ the total number of nodes in the system for $i=1 \dots N$.

\section{Minimizing the data download time}
\label{min_problem}

\begin{figure}[!h]
\centering
\vspace{-0.3cm}
\includegraphics[width=0.65\linewidth]{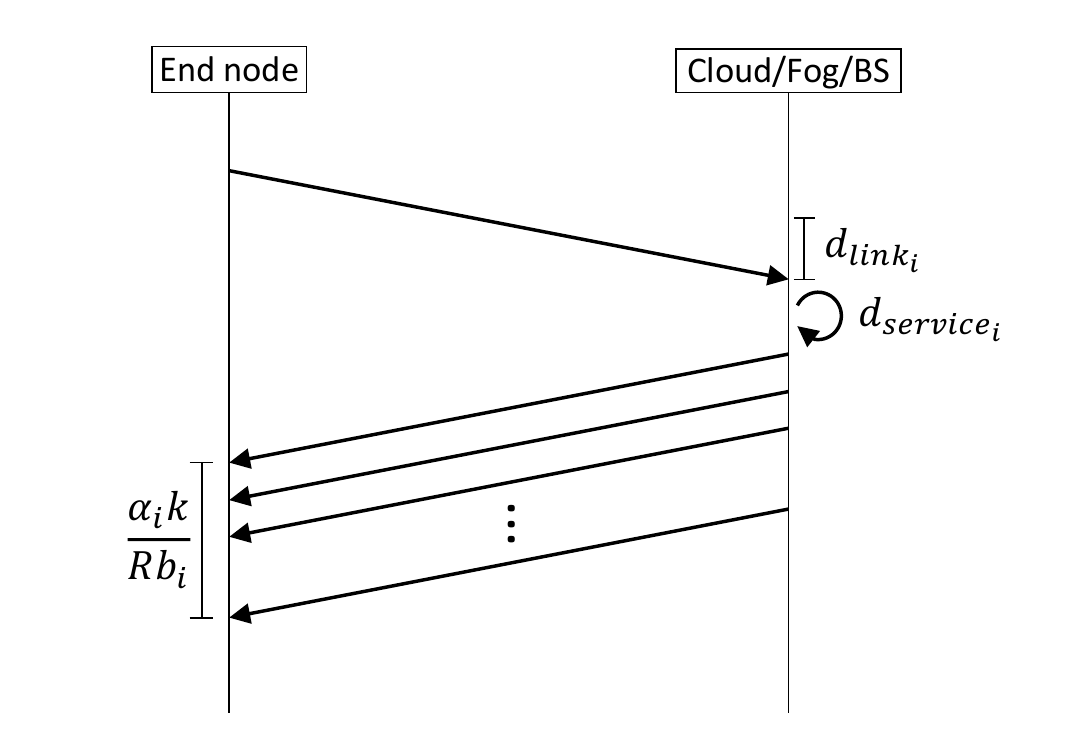}
\hspace{-1cm}
\caption{Download time estimation.}
\label{fig:downtime_esquema}
\end{figure}

The way data is distributed becomes the key factor to decrease the total download time. Based on Fig.~\ref{fig:downtime_esquema}, we calculate the overall download time, $T_i^{download}$, as the time each node $i$ ($0 \leq i \leq N$) spends sending the coded packets that will eventually contribute to the whole download process, i.e. to the final information acquisition at the end node. We can thus define it as follows: 

\begin{equation}
    T^{download}_i = d^{\text{request}}_i + \frac{\alpha_i k}{Rb_i} 
\label{T_download}
\end{equation}

\noindent where $Rb$ represents the download rate, $d^{\text{request}}_i$corresponds to the time needed to make the request to the corresponding node, and $\alpha_i$ represents the percentage of data stored at (retrieved from) such node.

Hence, $d^{\text{request}}_i$ can be calculated as the sum of the communication delay, $d^{\text{link}}_i$, and the service time, $d^{service}_i$:

\begin{equation}
    d^{\text{request}}_i = d^{\text{link}}_i + d^{service}_i 
\end{equation}

Previous works \cite{Sipos14} also sought the minimization of the total download time. However, they only considered the transmission rate to establish $\alpha$, so the data distribution did not consider the communication delay nor the service time. The information percentage to be stored at each node would be thus estimated as:

\begin{equation}
    \alpha_i = \frac{Rb_i}{\sum_{i=1}^N Rb_i} 
\end{equation}

Opposed to that, we also consider other contributions to the overall download time, as was seen earlier, to find the optimum distribution strategy. We pose an optimization problem that seeks minimizing such time.

For the specific scenario presented on this paper, the cost function is the total system download time, $T_{total}$. The variables are the information percentages stored at (retrieved from) each node, which conform the $\boldsymbol{\alpha}$ vector, while the corresponding constraints are shown in Eq.~\ref{constraints}. The first one ensures that all required packets are received, while the second one limits the $\alpha$ variables to be within the $[0, 1]$ interval. 

Given the total number of packets $k$, the request time,
$d_{request}$, and the download rate, $Rb$, at every system node, the cost function to be minimized in this paper is the total system download time, $T_{total}$. Note that since the download process is performed simultaneously from all nodes (cloud and fog), $T_{total}$ is thus determined by the download time of the slowest node. Therefore,

\begin{equation}
    T_{total}(\alpha_1, \ldots, \alpha_N)=\max_{i\in \{1,\ldots, N\}}T^{download}_i(\alpha_i,Rb_i,d_i^{request},k) 
\end{equation}

Thus, the constrained minimization problem aims to find the $\{\alpha^*_i\}_{i=1}^N$ that minimize $T_{total}(\alpha_1,\ldots, \alpha_N)$, that is,

\begin{align}
    & &(\alpha^*_1,\ldots , \alpha^*_N)=&\arg \min_{\alpha_1,\ldots, \alpha_N}T_{total}(\alpha_1,\ldots, \alpha_N)=\nonumber\\
    & & &\arg\min_{\alpha_1,\ldots, \alpha_N}\left( \max_{i\in \{1,\ldots, N\}}T^{download}_i(\alpha_i,Rb_i,d_i^{request},k)\right)=\\
    & & &\arg\min_{\alpha_1,\ldots, \alpha_N}\left( \max_{i\in \{1,\ldots, N\}}\left(d_i^{request}+\frac{\alpha_ik}{Rb_i}\right)\right)\nonumber
\end{align}

\noindent where we have use the $T^{download}_i $ given in Eq.~\ref{T_download} under the constraints:

\begin{equation}
    \sum_{i=1}^N\alpha_i=1\;\;\mbox{and}\;\; 0\leq \alpha_i\leq 1,\;i\in \{1,\ldots, N\}
\label{constraints}
\end{equation}




Considering that all functions are linear and so affine, it is straightforward to establish the convexity of the cost function, using Jensen's inequality \cite{Boyd04}. Hence, a unique global minimum exists. The solution of the problem can be represented as the following equation system:
\begingroup
\allowdisplaybreaks
\begin{align}
& 1- \sum_{i=1}^{N}\lambda_i=0 \notag \\
& a_i\lambda_i-\lambda_{N+i}+\lambda_{2N+i}+\nu=0 \;\;  && \forall i = 1 \ldots N \notag\\
& \lambda_i (a_i\alpha_i+b_i-z)=0  && \forall i = 1 \ldots N \\
& -\lambda_{N+i}\alpha_i = 0\notag  && \forall i = 1 \ldots N \\
& \lambda_{2N+i}(\alpha_i-1) = 0\notag  && \forall i = 1 \ldots N\\
& \nu (\sum_{i=1}^{N}\alpha_i-1)=0\notag
\end{align}
\endgroup

\noindent where $\lambda_i > 0$. A detailed description of the problem solution is given in \ref{appendix}.

%% file: Results.tex
\section{Discussion of results}
\label{results}
This paper focuses on analyzing the download time from a multi-tier architecture (fog and cloud) targeted to distributed data storage. We have applied the proposed model in different scenarios, using various distribution methods, in order to understand the impact of modifying the system parameters, as well as to compare the proposed scheme with alternative strategies. The reference scenario that we have used to carry out the experiments is next described. 

The system is composed of 5 cloud nodes, 3 fog nodes, and 1 BS. We assume the user wants to download 100~MB. To establish the service delay of the nodes, we have applied an average service time, $t_s$, of 20~ms for the clouds and 50~ms for the fogs. In addition, the traffic load, $\rho$, follows a uniform distribution within the interval $[0.4-0.9]$ and $[0.2-0.7]$ for cloud and fog nodes, respectively. The values for the download rate, $Rb$, and communication latency, $d_{link}$, are based on 4G technology, as was previously mentioned. We have applied a $d_{link}$ between 30-100~ms for the communication between the end node and BSs (either having a fog node or not), and we have assumed twice this delay when the communication is with a cloud node. Finally, the $Rb$ used for every node is randomly selected within the interval 15-72~Mbps.

The optimization problem introduced in Section~\ref{min_problem} captures the characteristics of a multi-tier distributed storage system, with the goal of achieving the minimum data download time. In order to solve such problem we use Python, particularly, the \emph{scipy.optimize.linprog} module from the SciPy library\footnote{https://docs.scipy.org/doc/scipy/reference/optimize.linprog-simplex.html}, which allows to include the corresponding bounds and constraints.


\subsection{Single Vs. multiple node storage}
\begin{figure}[!h]
	\centering
    \input{Figures/single_multi_boxplot.tex}
	\caption{Data download time applying a single-node and two different multiple-node strategies.}
\label{fig:single_multi}
\end{figure}
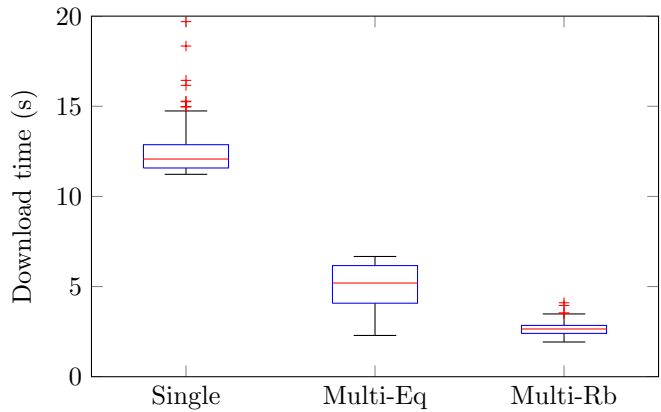

First, we have analyzed the time required to download the corresponding data over three different scenarios, refer to Fig.~\ref{fig:single_multi}. In the \emph{Single} configuration, the whole data is retrieved from the node having the best conditions. On the contrary, the other two scenarios correspond to distributed multi-node architectures, where data is shared among all the system nodes. In the \emph{Multi-Eq} scenario data is evenly stored in every node and so the terminal gets the same amount of information from all nodes, while in \emph{Multi-Rb}, the data distribution is performed based on the download rate of each node, as was proposed in \cite{Sipos14}. 

We have carried out experiments encompassing 200 independent snap-shots for the three configurations, where the traffic load at the nodes, $\rho$, and so $d_{service}$, randomly varies in each of them. As can be clearly seen in Fig.~\ref{fig:single_multi}, the use of multiple nodes yields much lower download times. Although in the first scenario the node from which data are downloaded corresponds to the one with the best conditions, when the distributed approach and so the capacity of all nodes is exploited, we can see a remarkable reduction on the download time. It is also important to note that when the distribution is made taking into account the particular conditions of the nodes, the download time can be further reduced. Furthermore, using multiple nodes to download the data leads to a more reliable system, since it would alleviate overload and poor connectivity situations.

\subsection{System parameter variation}
We have studied the overall download time varying the following system parameters: the number of nodes (both at cloud and fog level), the node load ($\rho$), and the generation size. 
We have used the two multi-node configurations that were previously described: (1) \emph{Eq}, where data are evenly distributed among all available nodes, and (2) \emph{Rb}, where the data distribution is planned considering the download rate of each node. In addition, we have also included our proposed scheme, which we refer to as (3) \emph{Opt}, since it is based on the previously described optimization problem. We thus calculate the amount of data to store in (retrieve from) each node considering, not only the download rate, but also the communication and service delays.

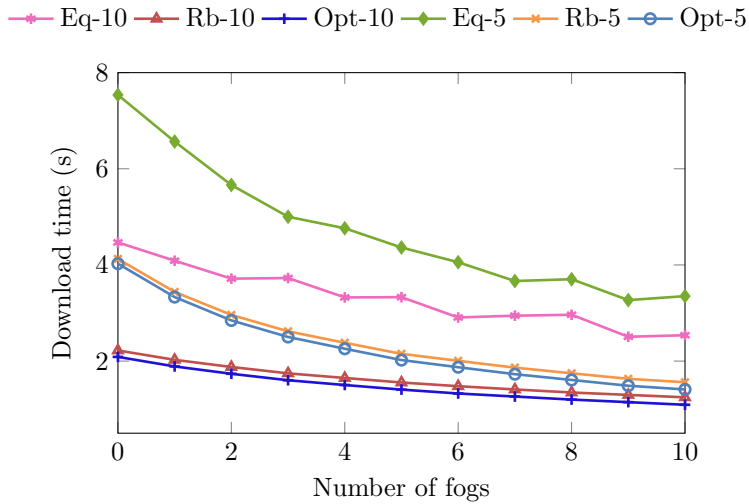
\begin{figure}[!h]
	\centering
 	\input{Figures/legend_nodes.tex} \break
 	\vspace{-5.8cm}
    \input{Figures/downtime_fogs.tex}
	\caption{Download time modifying the number of fog nodes in the system.}   
\label{fig:nodes}
\end{figure}

First, Fig.~\ref{fig:nodes} shows the results for the scenario where the available number of fogs in the system increases from 0 and 10, setting the number of cloud nodes to both 5 and 10. Conversely, the same analysis was performed varying the clouds, and fixing the number of fogs. The results were rather similar, so we only discuss the performance obtained when increasing the number of fog nodes. 

It can be observed that for the three distributed storage solutions, the download time decreases as more nodes are added to the system. When there are 10 clouds available, the overall download time is lower than when there are 5. However, in the latter case, the download time drops more sharply when increasing the number of fog nodes.  Moreover, we can see a clear difference between the \emph{Eq} method and the two other approaches. Furthermore, if we compare \emph{Opt} and \emph{Rb} schemes, it can be noted that the download time using the proposed solution is slightly lower. However, the benefit of the \emph{Opt} approach is not very relevant, since the communication delays are not very high, and stay within a rather short interval 30-100~ms. 


Next, we analyze how the download time is impacted when we vary the load of the available nodes. We have applied four different intervals for $\rho$: 0.1-0.3, 0.3-0.5, 0.5-0.7, and 0.7-0.9. In all cases, the load is randomly selected within the corresponding interval. When modifying the load of fog nodes, the load of the clouds is set to $[0.4-0.9]$. On the contrary, if we modify the cloud nodes' load, we fix $\rho$ between $[0.2-0.7]$ for the fog. 

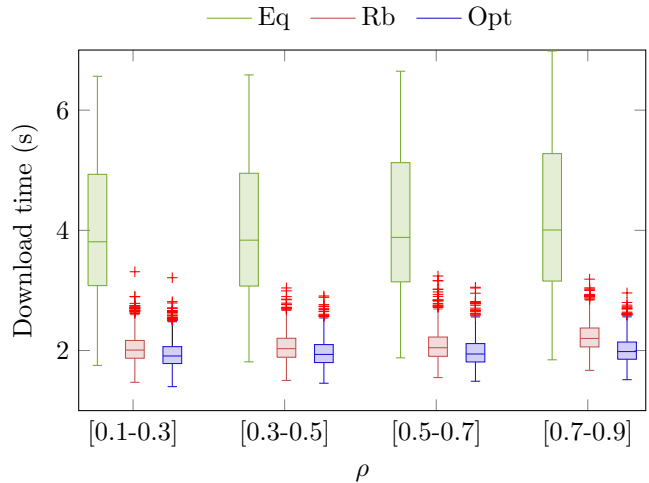
\begin{figure}[!h]
	\centering
    \input{Figures/rho_fog.tex} 
    \vspace{-0.2cm}
    \caption{Download time modifying the traffic load of fog nodes.}
\label{fig:rho_fog}
\end{figure}

Figure~\ref{fig:rho_fog}  shows  the  system  performance  when increasing the load of fog nodes. With the aforementioned configuration, we run 1000 independent experiments, and the figure shows the whisker plot of the observed download times for the three distribution methods, \emph{Eq}, \emph{Rb} and \emph{Opt}. In this case, the behavior of changing the load of either the cloud or the fog was rather alike, so we just show the results that were observed when increasing the load of the fog nodes, keeping $\rho$ within $[0.4-0.9]$ for the cloud nodes. First of all, we can see that the \emph{Eq} scheme, besides yielding a much slower performance, shows a rather unpredictable behavior, since the variance of the download time is quite high. The use of \emph{Rb} and \emph{Opt} does not only improve the overall behavior of the system, but it also helps to yield a more robust performance. We can see that the download time is not severely jeopardized when we increase the load of the fog nodes, and the corresponding variance is not very high. The figure also shows that the proposed scheme, which takes into account the load to establish the storage strategy, slightly outperforms the \emph{Rb} solution, yielding a smaller download time, as well as a more predictable behavior. 

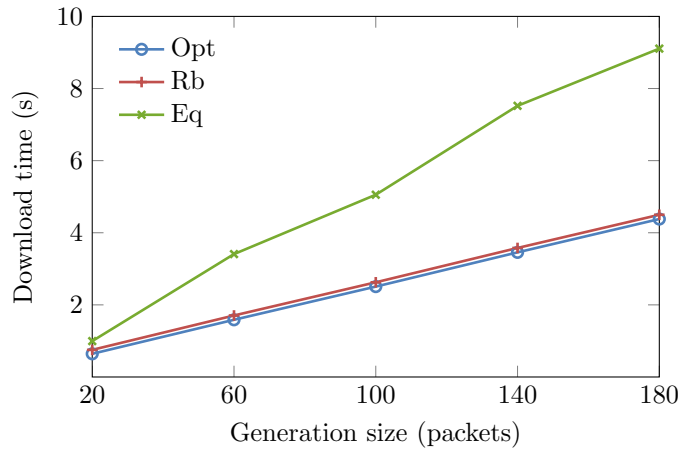
\begin{figure}[!h]
	\centering
    \input{Figures/generation_size.tex}
	\caption{Download time modifying the generation size.}
\label{fig:gen_size}
\end{figure}

The last parameter we have modified is the generation size, i.e. number of packets and so the amount of data to be stored in the system. We wanted to assess whether this had an impact on the system performance. Figure~\ref{fig:gen_size} shows how the average download time changes when increasing the generation size. We have used the configuration that was depicted at the beginning of the section, and we show the results for the three schemes we are analyzing. Both \emph{Rb} and \emph{Opt} clearly outperforms \emph{Eq}, especially when the amount of data to be retrieved increases. In the three cases there is an almost linear increasing trend, but the slope exhibited by the \emph{Eq} approach is greater. We also see that the proposed scheme (\emph{Opt}) slightly reduces the time that was observed for the \emph{Rb} solution.

\subsection{High-delay cases}
The results that have been discussed up to now are obtained with regular circumstances i.e. communication latency and service delays within reasonable values. However, it might happen that, due to network congestion or connectivity issues, end users are not able to communicate with either fog devices or cloud servers. Furthermore, some of those nodes could receive multiple requests, leading to overload or even outage situations. These situations would certainly cause an increase of communication and service delays, respectively. Since the proposed solution in this paper does not only consider \emph{Rb} to establish the distributed storage strategy, but it also takes into account such delays, it is under this `extreme' situations where we might actually see more clearly its benefits compared to the \emph{Rb} approach. Hence, using the reference scenario described at the beginning of this section, we have carried out multiple experiments applying both higher $d_{link}$ and $\rho$ values, in order to observe the behavior of the different distribution models. 

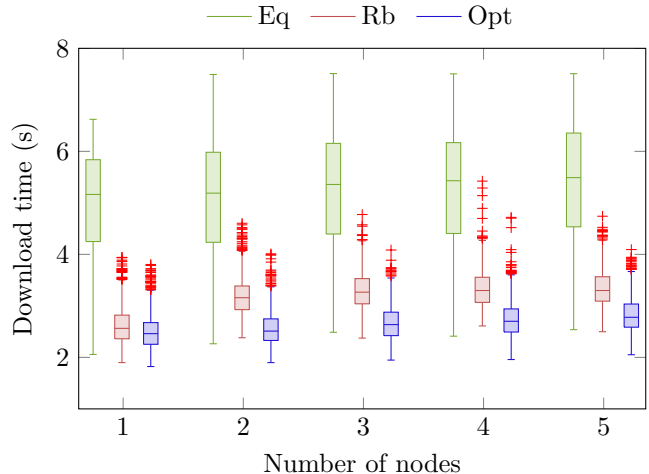
\begin{figure}[!h]
	\centering
    \input{Figures/dlink_boxplot.tex} 
    \vspace{-0.2cm}
    \caption{Download time increasing the number of nodes having high latency.}
\label{fig:high_latency}
\end{figure}

On a first setup, we have strongly increased the communication latency for various nodes, to be within the 0.5-1~s interval. We increase the number of nodes suffering from such connectivity issues, and we measure the download time for the three schemes we are studying. The obtained results are illustrated in Fig.~\ref{fig:high_latency}. Although we can see that the data download time increases when more nodes suffer from high communication delays, the \emph{Opt} approach is able to keep a comparable performance, and the increase is much less noticeable. As in previous experiments, the \emph{Eq} method yields much higher download times than the two other schemes. On the other hand, we can see in this case that the difference between \emph{Rb} and \emph{Opt} is more clear, showing that our proposal responds better to high communication latency values. 

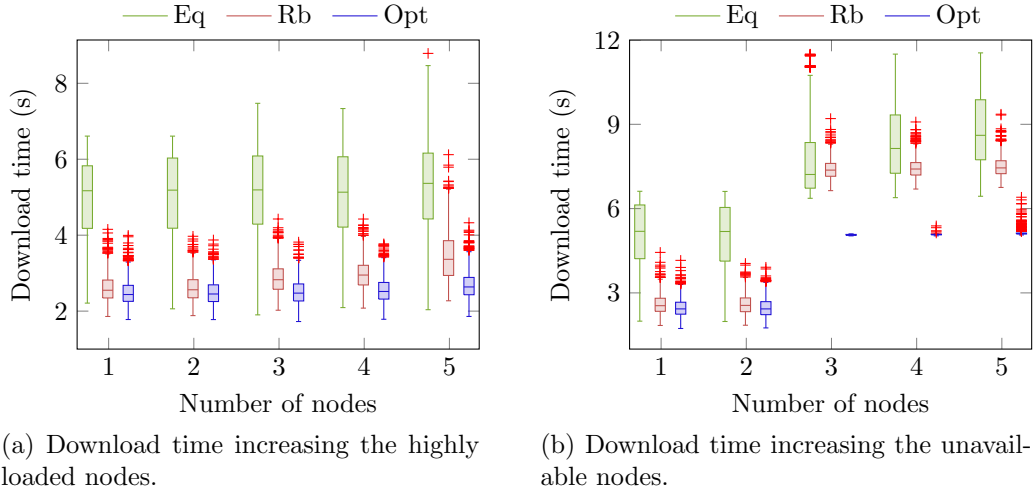
\begin{figure}[!h]
    \centering
    \subfloat[Download time increasing the highly loaded nodes.\label{fig:high_load}]{\input{Figures/load_boxplot.tex}}\hfill
    \subfloat[Download time increasing the unavailable nodes.\label{fig:overload}]{\input{Figures/overload_boxplot.tex}}
\caption{Download time increasing the number of (a) highly loaded and (b) unavailable nodes in the system.}   
\label{fig:high_loads}
\end{figure}

As has been already mentioned, $d_{request}$ is not only affected by the communication latency, but also by the time it takes to respond to the data request, $d_{service}$, which depends on the nodes' load level. Following a similar procedure as in the previous experiment, we assume that some of the nodes are rather overloaded, $\rho$ within the $[0.8-0.95]$ interval, and we increase the number of nodes that are currently having such condition. Fig.~\ref{fig:high_load} shows the download times obtained for the three schemes when increasing the number of highly loaded nodes in the system, from 1 to 5. As can be seen, when there are more overloaded nodes, we can see a clear increase of the download time for the three distribution methods. However, it has much less impact when using the \emph{Opt} scheme, which takes into account load nodes to establish the data distribution share at each node. In fact, the proposed scheme is able to keep the download time at the same level, showing its reliability. Furthermore, the difference between \emph{Rb} and \emph{Opt} can be also clearly seen, in particular when the number of overloaded nodes is 5. It is also worth noting that another advantage of the \emph{Opt} scheme is that it yields a much more predictable performance, since the whisker plots yield a lower variance than for the \emph{Rb} solution. 

Going a step further, we have also considered a situation in which the overloaded situation might actually imply an outage of some of the nodes. For that, we have mimicked the previous experiment, but in this case the load of the saturated nodes is fixed at $\rho = 0.99$, which almost reflects an unavailability circumstance. The obtained results are depicted in Fig.~\ref{fig:overload}. As could have been expected, the impact of increasing nodes in outage is quite strong for the three distribution methods. However, we can actually see more clearly the benefits of the \emph{Opt} scheme, which yields much lower download times than \emph{Rb}, showing as well a more predictable performance. In fact, in this particular situation, the performance of \emph{Rb} is, in average, comparable to that exhibited by the \emph{Eq} naive approach. 



Finally, Table~\ref{table:alphas} shows an illustrative example of how the three schemes would actually configure the data distribution among the system nodes for the experiment shown in Fig.~\ref{fig:overload}, for $\rho=0.99$ and $N=3$. It can be observed that the three solutions establish different $\alpha$ values for each node, reflecting the parameters they use to establish such strategy. For instance, when applying the \emph{Opt} method, it considers that one of the nodes (3) is far too slow and it would be therefore better not to retrieve data from that node. In this case, considering the impact of the corresponding delay, the difference between the used methods is quite relevant, being the proposed solution 2.8~s and 4.8~s faster than \emph{Rb} and \emph{Eq}, respectively. 

\begin{table}[!h]
	\centering
	\caption{$\alpha$ and $T_{total}$ values for each of the distribution methods.}
	\resizebox{0.95\linewidth}{!}{\begin{tabular}{|| >{\centering\arraybackslash} m{2.6cm} | c | c | c | c | c | c | c | c | c ||}
		\hline
        \textbf{Distribution method} & $\boldsymbol{\alpha_1}$ & $\boldsymbol{\alpha_2}$ & $\boldsymbol{\alpha_3}$ & $\boldsymbol{\alpha_4}$ & $\boldsymbol{\alpha_5}$ & $\boldsymbol{\alpha_6}$ & $\boldsymbol{\alpha_7}$ & $\boldsymbol{\alpha_8}$ & $\boldsymbol{T_{total} (s)}$\\ 
        \hline\hline
        \textbf{Eq} & 0.125 & 0.125 & 0.125 & 0.125 & 0.125 & 0.125 & 0.125 & 0.125 & 9.847\\ 
        \hline
        \textbf{Rb} & 0.257 & 0.094 & 0.0725 & 0.134 & 0.091 & 0.078 & 0.112 & 0.159 & 7.847\\ 
        \hline
        \textbf{Opt} & 0.149 & 0.119 & 0 & 0.173 & 0.112 & 0.098 & 0.142 & 0.206 & 5.047\\ 
        \hline
	\end{tabular}}	
	\label{table:alphas}
\end{table}

%% file: Figures/single_multi_boxplot.tex
%
%
\begin{tikzpicture}
font = \footnotesize, 
\begin{axis}[%
width=0.55\columnwidth,
height=0.35\columnwidth,
at={(0.758in,0.481in)},
scale only axis,
unbounded coords=jump,
xmin=0.5,
xmax=3.5,
xtick={1,2,3},
xticklabels={{Single},{Multi-Eq},{Multi-Rb}},
ymin=0,
ymax=20,
ylabel style={font=\color{white!15!black}},
ylabel={Download time (s)},
compat=1.3,
axis background/.style={fill=white},
legend style={legend cell align=left, align=left, draw=white!15!black},
]
\addplot [color=black, forget plot]
  table[row sep=crcr]{%
1	12.8688827174977\\
1	14.7405099010126\\
};
\addplot [color=black, forget plot]
  table[row sep=crcr]{%
2	6.16376855421874\\
2	6.66907897422688\\
};
\addplot [color=black, forget plot]
  table[row sep=crcr]{%
3	2.84496779891255\\
3	3.48264555464434\\
};
\addplot [color=black, forget plot]
  table[row sep=crcr]{%
1	11.2260944970777\\
1	11.5774895671846\\
};
\addplot [color=black, forget plot]
  table[row sep=crcr]{%
2	2.28851352511381\\
2	4.07780008129458\\
};
\addplot [color=black, forget plot]
  table[row sep=crcr]{%
3	1.92546053763316\\
3	2.40333907495016\\
};
\addplot [color=black, forget plot]
  table[row sep=crcr]{%
0.8875	14.7405099010126\\
1.1125	14.7405099010126\\
};
\addplot [color=black, forget plot]
  table[row sep=crcr]{%
1.8875	6.66907897422688\\
2.1125	6.66907897422688\\
};
\addplot [color=black, forget plot]
  table[row sep=crcr]{%
2.8875	3.48264555464434\\
3.1125	3.48264555464434\\
};
\addplot [color=black, forget plot]
  table[row sep=crcr]{%
0.8875	11.2260944970777\\
1.1125	11.2260944970777\\
};
\addplot [color=black, forget plot]
  table[row sep=crcr]{%
1.8875	2.28851352511381\\
2.1125	2.28851352511381\\
};
\addplot [color=black, forget plot]
  table[row sep=crcr]{%
2.8875	1.92546053763316\\
3.1125	1.92546053763316\\
};
\addplot [color=blue, forget plot]
  table[row sep=crcr]{%
0.775	11.5774895671846\\
0.775	12.8688827174977\\
1.225	12.8688827174977\\
1.225	11.5774895671846\\
0.775	11.5774895671846\\
};
\addplot [color=blue, forget plot]
  table[row sep=crcr]{%
1.775	4.07780008129458\\
1.775	6.16376855421874\\
2.225	6.16376855421874\\
2.225	4.07780008129458\\
1.775	4.07780008129458\\
};
\addplot [color=blue, forget plot]
  table[row sep=crcr]{%
2.775	2.40333907495016\\
2.775	2.84496779891255\\
3.225	2.84496779891255\\
3.225	2.40333907495016\\
2.775	2.40333907495016\\
};
\addplot [color=red, forget plot]
  table[row sep=crcr]{%
0.775	12.0728397255671\\
1.225	12.0728397255671\\
};
\addplot [color=red, forget plot]
  table[row sep=crcr]{%
1.775	5.19579307658285\\
2.225	5.19579307658285\\
};
\addplot [color=red, forget plot]
  table[row sep=crcr]{%
2.775	2.64784507981866\\
3.225	2.64784507981866\\
};
\addplot [color=black, draw=none, mark=+, mark options={solid, red}, forget plot]
  table[row sep=crcr]{%
1	14.9750295979407\\
1	14.9752209956887\\
1	14.9813021307552\\
1	15.252229740607\\
1	15.2796134268352\\
1	16.1606285460337\\
1	16.4369831742381\\
1	18.3410822843214\\
1	19.6952412276402\\
};
\addplot [color=black, draw=none, mark=+, mark options={solid, red}, forget plot]
  table[row sep=crcr]{%
nan	nan\\
};
\addplot [color=black, draw=none, mark=+, mark options={solid, red}, forget plot]
  table[row sep=crcr]{%
3	3.52409172090454\\
3	3.53325272370885\\
3	3.9533029247261\\
3	4.10244764160978\\
};
\end{axis}
\end{tikzpicture}%

%% file: Figures/legend_nodes.tex

\begin{tikzpicture} 
\input{doc/colors.tex}
font = \footnotesize,
    \begin{axis}[%
    hide axis,
    xmin=0,
    xmax=50,
    ymin=0,
    ymax=0.1,
    legend style={{anchor=west},at={(0.55, 1.15)}, anchor=north, fill=white, draw=none, legend columns=-1, font = \footnotesize}]
    
    \addlegendimage{clr11, solid,line width=1pt, mark=asterisk}
    \addlegendentry{Eq-10};
    
    \addlegendimage{clr4, solid,line width=1pt, mark=triangle}
    \addlegendentry{Rb-10};
    
    \addlegendimage{clr3, solid,line width=1pt, mark=+}
    \addlegendentry{Opt-10};
    
    \addlegendimage{clr9, solid,line width=1pt, mark=diamond*}
    \addlegendentry{Eq-5};
    
    \addlegendimage{clr2, solid,line width=1pt, mark=x}
    \addlegendentry{Rb-5};
    
    \addlegendimage{clr1, solid,line width=1pt, mark=o}
    \addlegendentry{Opt-5};
    
    \end{axis}
\end{tikzpicture}


%% file: Figures/downtime_fogs.tex
\begin{tikzpicture} 
\input{doc/colors.tex}
font = \footnotesize, 
\begin{axis}[legend pos=south west, scale only axis, 
yticklabel style={
      /pgf/number format/fixed,
      /pgf/number format/precision=5
},
scaled y ticks=false, 
width=0.55\columnwidth,
height=0.35\columnwidth,
mark options={solid},
ymin=0.5,
ymax=8,
xmin=0,
xmax=10,
ylabel={Download time (s)},
xlabel={Number of fogs},
compat=1.3,
legend style={/tikz/every even column/.append style={column sep=0.2cm}},
legend style={draw=none, fill=none,legend columns=1, font = \scriptsize},
legend cell align={left},
legend pos =north east,
]

\pgfplotstableread{Figures/data/downtime_fogs.dat}{\dataset}

\addplot [clr11, solid,line width=1pt, mark=asterisk]
table[x expr=\thisrowno{0}, y expr=\thisrowno{1}] from \dataset;
\addlegendentry{Eq-10 C}

\addplot [clr4, solid,line width=1pt, mark=triangle]
table[x expr=\thisrowno{0}, y expr=\thisrowno{2}] from \dataset;
\addlegendentry{Rb-10 C}

\addplot [clr3, solid,line width=1pt, mark=+]
table[x expr=\thisrowno{0}, y expr=\thisrowno{3}] from \dataset;
\addlegendentry{Opt-10 C}

\addplot [clr9, solid,line width=1pt, mark=diamond*]
table[x expr=\thisrowno{0}, y expr=\thisrowno{4}] from \dataset;
\addlegendentry{Eq-5 C}

\addplot [clr2, solid,line width=1pt, mark=x]
table[x expr=\thisrowno{0}, y expr=\thisrowno{5}] from \dataset;
\addlegendentry{Rb-5 C}

\addplot [clr1, solid,line width=1pt, mark=o]
plot[]
table[x expr=\thisrowno{0}, y expr=\thisrowno{6}] from \dataset;
\addlegendentry{Opt-5 C}

\legend{}

\end{axis}
\end{tikzpicture}

%% file: Figures/rho_fog.tex
\begin{tikzpicture} 
\input{doc/colors.tex}
font = \footnotesize, 
\begin{axis}[legend pos=south west, scale only axis, 
yticklabel style={
      /pgf/number format/fixed,
      /pgf/number format/precision=5
},
scaled y ticks=false, 
width=0.55\columnwidth,
height=0.35\columnwidth,
mark options={solid},
xmin=-0.0499999999999998,
xmax=31.75,
xtick={3,11.5,20,28.5},
xticklabels={{[0.1-0.3]},{[0.3-0.5]},{[0.5-0.7]},{[0.7-0.9]}},
xlabel style={font=\color{white!15!black}},
xlabel={$\rho$},
ymin=1,
ymax=7,
ylabel style={font=\color{white!15!black}},
ylabel={Download time (s)},
compat=1.3,
legend style={/tikz/every even column/.append style={column sep=0.2cm}},
legend style={{anchor=west},at={(0.5, 1.15)}, anchor=north, fill=white, draw=none, legend columns=-1, font = \footnotesize},
every axis plot/.append style={fill,fill opacity=0.2}
]

\addplot [color=clr9, forget plot]
  table[row sep=crcr]{%
1	4.9315840906067\\
1	6.56373608705301\\
};
\addplot [color=clr4, forget plot]
  table[row sep=crcr]{%
3.1	2.16744819319753\\
3.1	2.59701037022583\\
};
\addplot [color=black, forget plot]
  table[row sep=crcr]{%
5.2	2.0646011610915\\
5.2	2.47884950777735\\
};
\addplot [color=clr9, forget plot]
  table[row sep=crcr]{%
9.5	4.94966008698917\\
9.5	6.5860962096869\\
};
\addplot [color=clr4, forget plot]
  table[row sep=crcr]{%
11.6	2.20346036301101\\
11.6	2.66715394286805\\
};
\addplot [color=clr3, forget plot]
  table[row sep=crcr]{%
13.7	2.09906405332825\\
13.7	2.54283342897763\\
};
\addplot [color=clr9, forget plot]
  table[row sep=crcr]{%
18	5.12626693135986\\
18	6.64749416425282\\
};
\addplot [color=clr4, forget plot]
  table[row sep=crcr]{%
20.1	2.224323053831\\
20.1	2.69152263076307\\
};
\addplot [color=clr3, forget plot]
  table[row sep=crcr]{%
22.2	2.11592320590774\\
22.2	2.5593624162844\\
};
\addplot [color=clr9, forget plot]
  table[row sep=crcr]{%
26.5	5.27654484750623\\
26.5	6.98395731332813\\
};
\addplot [color=clr4, forget plot]
  table[row sep=crcr]{%
28.6	2.37470736262017\\
28.6	2.83827045316781\\
};
\addplot [color=clr3, forget plot]
  table[row sep=crcr]{%
30.7	2.14130648450359\\
30.7	2.56021678023489\\
};
\addplot [color=clr9, forget plot]
  table[row sep=crcr]{%
1	1.75195481479795\\
1	3.08024016492283\\
};
\addplot [color=clr4, forget plot]
  table[row sep=crcr]{%
3.1	1.4710375297013\\
3.1	1.87161121321169\\
};
\addplot [color=clr3, forget plot]
  table[row sep=crcr]{%
5.2	1.39897085928453\\
5.2	1.78354255082699\\
};
\addplot [color=clr9, forget plot]
  table[row sep=crcr]{%
9.5	1.81157917925492\\
9.5	3.07283254278631\\
};
\addplot [color=clr4, forget plot]
  table[row sep=crcr]{%
11.6	1.50247102116061\\
11.6	1.88853040651913\\
};
\addplot [color=clr3, forget plot]
  table[row sep=crcr]{%
13.7	1.45582611209056\\
13.7	1.7996995305929\\
};
\addplot [color=clr9, forget plot]
  table[row sep=crcr]{%
18	1.87830262784651\\
18	3.14274753759998\\
};
\addplot [color=clr4, forget plot]
  table[row sep=crcr]{%
20.1	1.54923216871192\\
20.1	1.90321913488366\\
};
\addplot [color=clr3, forget plot]
  table[row sep=crcr]{%
22.2	1.48862587077447\\
22.2	1.80972921852331\\
};
\addplot [color=clr9, forget plot]
  table[row sep=crcr]{%
26.5	1.84571665961616\\
26.5	3.15571904914076\\
};
\addplot [color=clr4, forget plot]
  table[row sep=crcr]{%
28.6	1.66962315112202\\
28.6	2.06020973785877\\
};
\addplot [color=clr3, forget plot]
  table[row sep=crcr]{%
30.7	1.51456823624472\\
30.7	1.85508214722162\\
};
\addplot [color=clr9, forget plot]
  table[row sep=crcr]{%
0.7375	6.56373608705301\\
1.2625	6.56373608705301\\
};
\addplot [color=clr4, forget plot]
  table[row sep=crcr]{%
2.8375	2.59701037022583\\
3.3625	2.59701037022583\\
};
\addplot [color=clr3, forget plot]
  table[row sep=crcr]{%
4.9375	2.47884950777735\\
5.4625	2.47884950777735\\
};
\addplot [color=clr9, forget plot]
  table[row sep=crcr]{%
9.2375	6.5860962096869\\
9.7625	6.5860962096869\\
};
\addplot [color=clr4, forget plot]
  table[row sep=crcr]{%
11.3375	2.66715394286805\\
11.8625	2.66715394286805\\
};
\addplot [color=clr3, forget plot]
  table[row sep=crcr]{%
13.4375	2.54283342897763\\
13.9625	2.54283342897763\\
};
\addplot [color=clr9, forget plot]
  table[row sep=crcr]{%
17.7375	6.64749416425282\\
18.2625	6.64749416425282\\
};
\addplot [color=clr4, forget plot]
  table[row sep=crcr]{%
19.8375	2.69152263076307\\
20.3625	2.69152263076307\\
};
\addplot [color=clr3, forget plot]
  table[row sep=crcr]{%
21.9375	2.5593624162844\\
22.4625	2.5593624162844\\
};
\addplot [color=clr9, forget plot]
  table[row sep=crcr]{%
26.2375	6.98395731332813\\
26.7625	6.98395731332813\\
};
\addplot [color=clr4, forget plot]
  table[row sep=crcr]{%
28.3375	2.83827045316781\\
28.8625	2.83827045316781\\
};
\addplot [color=clr3, forget plot]
  table[row sep=crcr]{%
30.4375	2.56021678023489\\
30.9625	2.56021678023489\\
};
\addplot [color=clr9, forget plot]
  table[row sep=crcr]{%
0.7375	1.75195481479795\\
1.2625	1.75195481479795\\
};
\addplot [color=clr4, forget plot]
  table[row sep=crcr]{%
2.8375	1.4710375297013\\
3.3625	1.4710375297013\\
};
\addplot [color=clr3, forget plot]
  table[row sep=crcr]{%
4.9375	1.39897085928453\\
5.4625	1.39897085928453\\
};
\addplot [color=clr9, forget plot]
  table[row sep=crcr]{%
9.2375	1.81157917925492\\
9.7625	1.81157917925492\\
};
\addplot [color=clr4, forget plot]
  table[row sep=crcr]{%
11.3375	1.50247102116061\\
11.8625	1.50247102116061\\
};
\addplot [color=clr3, forget plot]
  table[row sep=crcr]{%
13.4375	1.45582611209056\\
13.9625	1.45582611209056\\
};
\addplot [color=clr9, forget plot]
  table[row sep=crcr]{%
17.7375	1.87830262784651\\
18.2625	1.87830262784651\\
};
\addplot [color=clr4, forget plot]
  table[row sep=crcr]{%
19.8375	1.54923216871192\\
20.3625	1.54923216871192\\
};
\addplot [color=clr3, forget plot]
  table[row sep=crcr]{%
21.9375	1.48862587077447\\
22.4625	1.48862587077447\\
};
\addplot [color=clr9, forget plot]
  table[row sep=crcr]{%
26.2375	1.84571665961616\\
26.7625	1.84571665961616\\
};
\addplot [color=clr4, forget plot]
  table[row sep=crcr]{%
28.3375	1.66962315112202\\
28.8625	1.66962315112202\\
};
\addplot [color=clr3, forget plot]
  table[row sep=crcr]{%
30.4375	1.51456823624472\\
30.9625	1.51456823624472\\
};
\addplot [color=clr9, forget plot]
  table[row sep=crcr]{%
0.475	3.08024016492283\\
0.475	4.9315840906067\\
1.525	4.9315840906067\\
1.525	3.08024016492283\\
0.475	3.08024016492283\\
};
\addplot [color=clr4, forget plot]
  table[row sep=crcr]{%
2.575	1.87161121321169\\
2.575	2.16744819319753\\
3.625	2.16744819319753\\
3.625	1.87161121321169\\
2.575	1.87161121321169\\
};
\addplot [color=clr3, forget plot]
  table[row sep=crcr]{%
4.675	1.78354255082699\\
4.675	2.0646011610915\\
5.725	2.0646011610915\\
5.725	1.78354255082699\\
4.675	1.78354255082699\\
};
\addplot [color=clr9, forget plot]
  table[row sep=crcr]{%
8.975	3.07283254278631\\
8.975	4.94966008698917\\
10.025	4.94966008698917\\
10.025	3.07283254278631\\
8.975	3.07283254278631\\
};
\addplot [color=clr4, forget plot]
  table[row sep=crcr]{%
11.075	1.88853040651913\\
11.075	2.20346036301101\\
12.125	2.20346036301101\\
12.125	1.88853040651913\\
11.075	1.88853040651913\\
};
\addplot [color=clr3, forget plot]
  table[row sep=crcr]{%
13.175	1.7996995305929\\
13.175	2.09906405332825\\
14.225	2.09906405332825\\
14.225	1.7996995305929\\
13.175	1.7996995305929\\
};
\addplot [color=clr9, forget plot]
  table[row sep=crcr]{%
17.475	3.14274753759998\\
17.475	5.12626693135986\\
18.525	5.12626693135986\\
18.525	3.14274753759998\\
17.475	3.14274753759998\\
};
\addplot [color=clr4, forget plot]
  table[row sep=crcr]{%
19.575	1.90321913488366\\
19.575	2.224323053831\\
20.625	2.224323053831\\
20.625	1.90321913488366\\
19.575	1.90321913488366\\
};
\addplot [color=clr3, forget plot]
  table[row sep=crcr]{%
21.675	1.80972921852331\\
21.675	2.11592320590774\\
22.725	2.11592320590774\\
22.725	1.80972921852331\\
21.675	1.80972921852331\\
};
\addplot [color=clr9, forget plot]
  table[row sep=crcr]{%
25.975	3.15571904914076\\
25.975	5.27654484750623\\
27.025	5.27654484750623\\
27.025	3.15571904914076\\
25.975	3.15571904914076\\
};
\addplot [color=clr4, forget plot]
  table[row sep=crcr]{%
28.075	2.06020973785877\\
28.075	2.37470736262017\\
29.125	2.37470736262017\\
29.125	2.06020973785877\\
28.075	2.06020973785877\\
};
\addplot [color=clr3, forget plot]
  table[row sep=crcr]{%
30.175	1.85508214722162\\
30.175	2.14130648450359\\
31.225	2.14130648450359\\
31.225	1.85508214722162\\
30.175	1.85508214722162\\
};
\addplot [color=clr9, forget plot]
  table[row sep=crcr]{%
0.475	3.80953512781102\\
1.525	3.80953512781102\\
};
\addplot [color=clr4, forget plot]
  table[row sep=crcr]{%
2.575	2.00755183694799\\
3.625	2.00755183694799\\
};
\addplot [color=clr3, forget plot]
  table[row sep=crcr]{%
4.675	1.90938586419248\\
5.725	1.90938586419248\\
};
\addplot [color=clr9, forget plot]
  table[row sep=crcr]{%
8.975	3.83650800661232\\
10.025	3.83650800661232\\
};
\addplot [color=clr4, forget plot]
  table[row sep=crcr]{%
11.075	2.03096734808616\\
12.125	2.03096734808616\\
};
\addplot [color=clr3, forget plot]
  table[row sep=crcr]{%
13.175	1.93484734813419\\
14.225	1.93484734813419\\
};
\addplot [color=clr9, forget plot]
  table[row sep=crcr]{%
17.475	3.882446151579\\
18.525	3.882446151579\\
};
\addplot [color=clr4, forget plot]
  table[row sep=crcr]{%
19.575	2.04665652523255\\
20.625	2.04665652523255\\
};
\addplot [color=clr3, forget plot]
  table[row sep=crcr]{%
21.675	1.94225233262716\\
22.725	1.94225233262716\\
};
\addplot [color=clr9]
  table[row sep=crcr]{%
25.975	4.00549597778778\\
27.025	4.00549597778778\\
};
\addlegendentry{Eq}

\addplot [color=clr4]
  table[row sep=crcr]{%
28.075	2.20102258106751\\
29.125	2.20102258106751\\
};
\addlegendentry{Rb}

\addplot [color=clr3]
  table[row sep=crcr]{%
30.175	1.98300058384113\\
31.225	1.98300058384113\\
};
\addlegendentry{Opt}

\addplot [color=black, draw=none, mark=+, mark options={solid, red}, forget plot]
  table[row sep=crcr]{%
nan	nan\\
};
\addplot [color=black, draw=none, mark=+, mark options={solid, red}, forget plot]
  table[row sep=crcr]{%
3.1	2.61584371105374\\
3.1	2.63881976170355\\
3.1	2.64238281901595\\
3.1	2.64964788114868\\
3.1	2.65947239981904\\
3.1	2.66236873342578\\
3.1	2.66818842899209\\
3.1	2.66850335910937\\
3.1	2.66969800663596\\
3.1	2.67369950166043\\
3.1	2.69769854060695\\
3.1	2.69928136612673\\
3.1	2.7003937163939\\
3.1	2.70040907671234\\
3.1	2.70240817440756\\
3.1	2.70570556943573\\
3.1	2.71463290796272\\
3.1	2.71584000838096\\
3.1	2.72205410317838\\
3.1	2.72830278439538\\
3.1	2.74549910215307\\
3.1	2.74717915818316\\
3.1	2.74926071672792\\
3.1	2.78427738141897\\
3.1	2.89601103293069\\
3.1	2.89790515837139\\
3.1	3.31088898147284\\
};
\addplot [color=black, draw=none, mark=+, mark options={solid, red}, forget plot]
  table[row sep=crcr]{%
5.2	2.49717717375213\\
5.2	2.50269323680581\\
5.2	2.51297099362134\\
5.2	2.51528326286592\\
5.2	2.52611943391433\\
5.2	2.53271079736257\\
5.2	2.53688770086017\\
5.2	2.54059259557522\\
5.2	2.54841356461313\\
5.2	2.5507936636515\\
5.2	2.55244430669304\\
5.2	2.57556925888176\\
5.2	2.58488464641521\\
5.2	2.6086310792716\\
5.2	2.63268689874746\\
5.2	2.63543654637203\\
5.2	2.64001953093655\\
5.2	2.65162342790709\\
5.2	2.65632269392703\\
5.2	2.66583405238004\\
5.2	2.70156230309774\\
5.2	2.71045255031906\\
5.2	2.79103254469053\\
5.2	2.81364680426328\\
5.2	3.21142249736453\\
};
\addplot [color=black, draw=none, mark=+, mark options={solid, red}, forget plot]
  table[row sep=crcr]{%
nan	nan\\
};
\addplot [color=black, draw=none, mark=+, mark options={solid, red}, forget plot]
  table[row sep=crcr]{%
11.6	2.67817863237294\\
11.6	2.69154617574784\\
11.6	2.70562340948014\\
11.6	2.71096356997787\\
11.6	2.71798399587643\\
11.6	2.71798967279373\\
11.6	2.76383534233885\\
11.6	2.76641578499586\\
11.6	2.7879296556633\\
11.6	2.845393105167\\
11.6	2.85450876375628\\
11.6	2.89728536668255\\
11.6	2.99260816319085\\
11.6	3.0413816803862\\
};
\addplot [color=black, draw=none, mark=+, mark options={solid, red}, forget plot]
  table[row sep=crcr]{%
13.7	2.5673096380547\\
13.7	2.57114652850578\\
13.7	2.58044912846205\\
13.7	2.58913580697638\\
13.7	2.60348296485123\\
13.7	2.64817727612006\\
13.7	2.65500436671524\\
13.7	2.68413160611987\\
13.7	2.70269129116904\\
13.7	2.73510816230348\\
13.7	2.73677936449677\\
13.7	2.76396519632638\\
13.7	2.88477935057613\\
13.7	2.90761222492005\\
};
\addplot [color=black, draw=none, mark=+, mark options={solid, red}, forget plot]
  table[row sep=crcr]{%
nan	nan\\
};
\addplot [color=black, draw=none, mark=+, mark options={solid, red}, forget plot]
  table[row sep=crcr]{%
20.1	2.71712576634016\\
20.1	2.72763873858523\\
20.1	2.74197637923678\\
20.1	2.74252319009811\\
20.1	2.75815915891746\\
20.1	2.7836003689025\\
20.1	2.82565384150059\\
20.1	2.82670295166126\\
20.1	2.83829198348326\\
20.1	2.83926270265753\\
20.1	2.8400752507706\\
20.1	2.90073889113935\\
20.1	2.93436549443218\\
20.1	2.9751600801648\\
20.1	3.01911170752587\\
20.1	3.01971367114056\\
20.1	3.16128518650532\\
20.1	3.16271082319374\\
20.1	3.24026485144393\\
};
\addplot [color=black, draw=none, mark=+, mark options={solid, red}, forget plot]
  table[row sep=crcr]{%
22.2	2.57985409833204\\
22.2	2.60518513870152\\
22.2	2.60777895793466\\
22.2	2.6181087492382\\
22.2	2.64638331779744\\
22.2	2.6604405200589\\
22.2	2.66577890925871\\
22.2	2.6790337804208\\
22.2	2.70493969004378\\
22.2	2.752539723414\\
22.2	2.7780657719241\\
22.2	2.78223332748617\\
22.2	2.78790124095843\\
22.2	2.80249034593121\\
22.2	2.80916192460608\\
22.2	2.81514232924896\\
22.2	2.95357771385196\\
22.2	3.04034908227294\\
22.2	3.05273038981757\\
};
\addplot [color=black, draw=none, mark=+, mark options={solid, red}, forget plot]
  table[row sep=crcr]{%
nan	nan\\
};
\addplot [color=black, draw=none, mark=+, mark options={solid, red}, forget plot]
  table[row sep=crcr]{%
28.6	2.85178042537687\\
28.6	2.87281519376071\\
28.6	2.8937230265818\\
28.6	2.89398358789933\\
28.6	2.89893647809412\\
28.6	2.90173267112118\\
28.6	2.90805886096965\\
28.6	2.90899558840867\\
28.6	2.91261922283458\\
28.6	2.91330106465569\\
28.6	2.91636150977183\\
28.6	2.9183593664177\\
28.6	2.9237099381088\\
28.6	2.93386754370455\\
28.6	3.00137184467474\\
28.6	3.00927906661037\\
28.6	3.02524013427083\\
28.6	3.04315710191622\\
28.6	3.18736959537873\\
};
\addplot [color=black, draw=none, mark=+, mark options={solid, red}, forget plot]
  table[row sep=crcr]{%
30.7	2.58584996004532\\
30.7	2.59864054987622\\
30.7	2.59895131419158\\
30.7	2.59992256773798\\
30.7	2.60027547726378\\
30.7	2.61055744238966\\
30.7	2.61169366330253\\
30.7	2.6140997173172\\
30.7	2.62387778436069\\
30.7	2.63867920987748\\
30.7	2.69308240937888\\
30.7	2.69610363776892\\
30.7	2.69839019118435\\
30.7	2.72020498460908\\
30.7	2.72708257693483\\
30.7	2.75508291120519\\
30.7	2.79956276459114\\
30.7	2.95832711376823\\
};

\end{axis}
\end{tikzpicture}%

%% file: Figures/generation_size.tex
\begin{tikzpicture} 
\input{doc/colors.tex}
font = \footnotesize, 
\begin{axis}[legend pos=south west, scale only axis, 
yticklabel style={
      /pgf/number format/fixed,
      /pgf/number format/precision=5
},
scaled y ticks=false, 
width=0.55\columnwidth,
height=0.35\columnwidth,
mark options={solid},
ymin=0,
ymax=10,
xmin=20,
xmax=180,
ytick={2,4,6,8,10},
yticklabels={2,4,6,8,10},
xtick={20,60,100,140,180},
xticklabels={20,60,100,140,180},
ylabel={Download time (s)},
xlabel={Generation size (packets)},
compat=1.3,
legend style={draw=none, fill=none,legend columns=1, font = \footnotesize},
legend cell align={left},
legend pos =north west,
]

\pgfplotstableread{Figures/data/generation_size_all.dat}{\dataset} 

\addplot [clr1, solid,line width=1pt, mark=o]
plot[]
table[x expr=\thisrowno{0}, y expr=\thisrowno{1}] from \dataset;
\addlegendentry{Opt}

\addplot [clr4, solid,line width=1pt, mark=+]
table[x expr=\thisrowno{0}, y expr=\thisrowno{2}] from \dataset;
\addlegendentry{Rb}

\addplot [clr9, solid,line width=1pt, mark=x]
table[x expr=\thisrowno{0}, y expr=\thisrowno{3}] from \dataset;
\addlegendentry{Eq}

\end{axis}
\end{tikzpicture}

%% file: Figures/dlink_boxplot.tex
%
%
\input{doc/colors.tex}

\begin{tikzpicture}
font = \footnotesize,
\begin{axis}[%
width=0.55\columnwidth,
height=0.35\columnwidth,
scale only axis,
unbounded coords=jump,
xmin=-0.199999999999999,
xmax=47,
xtick={3.5,13.5,23.5,33.5,43.5},
xticklabels={{1},{2},{3},{4},{5}},
xlabel style={font=\color{white!15!black}},
xlabel={Number of nodes},
ymin=1,
ymax=8,
ylabel style={font=\color{white!15!black}},
ylabel={Download time (s)},
compat=1.3,
axis background/.style={fill=white},
legend style={/tikz/every even column/.append style={column sep=0.2cm}},
legend style={{anchor=west},at={(0.5, 1.15)}, anchor=north, fill=white, draw=none, legend columns=-1, font = \footnotesize},
every axis plot/.append style={fill,fill opacity=0.2}
]
\addplot [color=clr9,   forget plot]
  table[row sep=crcr]{%
1	5.83775791733067\\
1	6.62171392218357\\
};
\addplot [color=clr4,   forget plot]
  table[row sep=crcr]{%
3.4	2.82024662066533\\
3.4	3.50170046464745\\
};
\addplot [color=clr3,   forget plot]
  table[row sep=crcr]{%
5.8	2.67534991860609\\
5.8	3.30588370811117\\
};
\addplot [color=clr9,   forget plot]
  table[row sep=crcr]{%
11	5.98189257540055\\
11	7.49246016717121\\
};
\addplot [color=clr4,   forget plot]
  table[row sep=crcr]{%
13.4	3.38597023786733\\
13.4	4.0582045703684\\
};
\addplot [color=clr3,   forget plot]
  table[row sep=crcr]{%
15.8	2.74754824964853\\
15.8	3.36944426336048\\
};
\addplot [color=clr9,   forget plot]
  table[row sep=crcr]{%
21	6.15452938201017\\
21	7.50990193974074\\
};
\addplot [color=clr4,   forget plot]
  table[row sep=crcr]{%
23.4	3.52923784621939\\
23.4	4.26152843833795\\
};
\addplot [color=clr3,   forget plot]
  table[row sep=crcr]{%
25.8	2.87742626213708\\
25.8	3.54139806020313\\
};
\addplot [color=clr9,   forget plot]
  table[row sep=crcr]{%
31	6.16895525981445\\
31	7.50422647045433\\
};
\addplot [color=clr4,   forget plot]
  table[row sep=crcr]{%
33.4	3.55552571978686\\
33.4	4.27019586593192\\
};
\addplot [color=clr3,   forget plot]
  table[row sep=crcr]{%
35.8	2.94084294219502\\
35.8	3.60183264138788\\
};
\addplot [color=clr9,   forget plot]
  table[row sep=crcr]{%
41	6.35518602965177\\
41	7.5079690058379\\
};
\addplot [color=clr4,   forget plot]
  table[row sep=crcr]{%
43.4	3.56540933746392\\
43.4	4.27475958446534\\
};
\addplot [color=clr3,   forget plot]
  table[row sep=crcr]{%
45.8	3.03555441226688\\
45.8	3.66701107954972\\
};
\addplot [color=clr9,   forget plot]
  table[row sep=crcr]{%
1	2.05680266705166\\
1	4.24895351432168\\
};
\addplot [color=clr4,   forget plot]
  table[row sep=crcr]{%
3.4	1.89848835823777\\
3.4	2.36135732086697\\
};
\addplot [color=clr3,   forget plot]
  table[row sep=crcr]{%
5.8	1.82228671625153\\
5.8	2.25392230899059\\
};
\addplot [color=clr9,   forget plot]
  table[row sep=crcr]{%
11	2.26438906626606\\
11	4.2341853371067\\
};
\addplot [color=clr4,   forget plot]
  table[row sep=crcr]{%
13.4	2.38108425359521\\
13.4	2.92640747675317\\
};
\addplot [color=clr3,   forget plot]
  table[row sep=crcr]{%
15.8	1.89847717017804\\
15.8	2.32858418595226\\
};
\addplot [color=clr9,   forget plot]
  table[row sep=crcr]{%
21	2.4873555883222\\
21	4.3949120603055\\
};
\addplot [color=clr4,   forget plot]
  table[row sep=crcr]{%
23.4	2.37448502304016\\
23.4	3.03980573071465\\
};
\addplot [color=clr3,   forget plot]
  table[row sep=crcr]{%
25.8	1.94715781496652\\
25.8	2.42224997460902\\
};
\addplot [color=clr9,   forget plot]
  table[row sep=crcr]{%
31	2.41155311805032\\
31	4.40555252234057\\
};
\addplot [color=clr4,   forget plot]
  table[row sep=crcr]{%
33.4	2.61029507227863\\
33.4	3.06773317688859\\
};
\addplot [color=clr3,   forget plot]
  table[row sep=crcr]{%
35.8	1.95788649499451\\
35.8	2.49199858010791\\
};
\addplot [color=clr9,   forget plot]
  table[row sep=crcr]{%
41	2.53841741997811\\
41	4.53481749169849\\
};
\addplot [color=clr4,   forget plot]
  table[row sep=crcr]{%
43.4	2.49829081412189\\
43.4	3.09155458041362\\
};
\addplot [color=clr3,   forget plot]
  table[row sep=crcr]{%
45.8	2.05153933053276\\
45.8	2.58616117298037\\
};
\addplot [color=clr9, forget plot]
  table[row sep=crcr]{%
0.7	6.62171392218357\\
1.3	6.62171392218357\\
};
\addplot [color=clr4, forget plot]
  table[row sep=crcr]{%
3.1	3.50170046464745\\
3.7	3.50170046464745\\
};
\addplot [color=clr3, forget plot]
  table[row sep=crcr]{%
5.5	3.30588370811117\\
6.1	3.30588370811117\\
};
\addplot [color=clr9, forget plot]
  table[row sep=crcr]{%
10.7	7.49246016717121\\
11.3	7.49246016717121\\
};
\addplot [color=clr4, forget plot]
  table[row sep=crcr]{%
13.1	4.0582045703684\\
13.7	4.0582045703684\\
};
\addplot [color=clr3, forget plot]
  table[row sep=crcr]{%
15.5	3.36944426336048\\
16.1	3.36944426336048\\
};
\addplot [color=clr9, forget plot]
  table[row sep=crcr]{%
20.7	7.50990193974074\\
21.3	7.50990193974074\\
};
\addplot [color=clr4, forget plot]
  table[row sep=crcr]{%
23.1	4.26152843833795\\
23.7	4.26152843833795\\
};
\addplot [color=clr3, forget plot]
  table[row sep=crcr]{%
25.5	3.54139806020313\\
26.1	3.54139806020313\\
};
\addplot [color=clr9, forget plot]
  table[row sep=crcr]{%
30.7	7.50422647045433\\
31.3	7.50422647045433\\
};
\addplot [color=clr4, forget plot]
  table[row sep=crcr]{%
33.1	4.27019586593192\\
33.7	4.27019586593192\\
};
\addplot [color=clr3, forget plot]
  table[row sep=crcr]{%
35.5	3.60183264138788\\
36.1	3.60183264138788\\
};
\addplot [color=clr9, forget plot]
  table[row sep=crcr]{%
40.7	7.5079690058379\\
41.3	7.5079690058379\\
};
\addplot [color=clr4, forget plot]
  table[row sep=crcr]{%
43.1	4.27475958446534\\
43.7	4.27475958446534\\
};
\addplot [color=clr3, forget plot]
  table[row sep=crcr]{%
45.5	3.66701107954972\\
46.1	3.66701107954972\\
};
\addplot [color=clr9, forget plot]
  table[row sep=crcr]{%
0.7	2.05680266705166\\
1.3	2.05680266705166\\
};
\addplot [color=clr4, forget plot]
  table[row sep=crcr]{%
3.1	1.89848835823777\\
3.7	1.89848835823777\\
};
\addplot [color=clr3, forget plot]
  table[row sep=crcr]{%
5.5	1.82228671625153\\
6.1	1.82228671625153\\
};
\addplot [color=clr9, forget plot]
  table[row sep=crcr]{%
10.7	2.26438906626606\\
11.3	2.26438906626606\\
};
\addplot [color=clr4, forget plot]
  table[row sep=crcr]{%
13.1	2.38108425359521\\
13.7	2.38108425359521\\
};
\addplot [color=clr3, forget plot]
  table[row sep=crcr]{%
15.5	1.89847717017804\\
16.1	1.89847717017804\\
};
\addplot [color=clr9, forget plot]
  table[row sep=crcr]{%
20.7	2.4873555883222\\
21.3	2.4873555883222\\
};
\addplot [color=clr4, forget plot]
  table[row sep=crcr]{%
23.1	2.37448502304016\\
23.7	2.37448502304016\\
};
\addplot [color=clr3, forget plot]
  table[row sep=crcr]{%
25.5	1.94715781496652\\
26.1	1.94715781496652\\
};
\addplot [color=clr9, forget plot]
  table[row sep=crcr]{%
30.7	2.41155311805032\\
31.3	2.41155311805032\\
};
\addplot [color=clr4, forget plot]
  table[row sep=crcr]{%
33.1	2.61029507227863\\
33.7	2.61029507227863\\
};
\addplot [color=clr3, forget plot]
  table[row sep=crcr]{%
35.5	1.95788649499451\\
36.1	1.95788649499451\\
};
\addplot [color=clr9, forget plot]
  table[row sep=crcr]{%
40.7	2.53841741997811\\
41.3	2.53841741997811\\
};
\addplot [color=clr4, forget plot]
  table[row sep=crcr]{%
43.1	2.49829081412189\\
43.7	2.49829081412189\\
};
\addplot [color=clr3, forget plot]
  table[row sep=crcr]{%
45.5	2.05153933053276\\
46.1	2.05153933053276\\
};
\addplot [color=clr9, forget plot]
  table[row sep=crcr]{%
0.4	4.24895351432168\\
0.4	5.83775791733067\\
1.6	5.83775791733067\\
1.6	4.24895351432168\\
0.4	4.24895351432168\\
};
\addplot [color=clr4, forget plot]
  table[row sep=crcr]{%
2.8	2.36135732086697\\
2.8	2.82024662066533\\
4	2.82024662066533\\
4	2.36135732086697\\
2.8	2.36135732086697\\
};
\addplot [color=clr3, forget plot]
  table[row sep=crcr]{%
5.2	2.25392230899059\\
5.2	2.67534991860609\\
6.4	2.67534991860609\\
6.4	2.25392230899059\\
5.2	2.25392230899059\\
};
\addplot [color=clr9, forget plot]
  table[row sep=crcr]{%
10.4	4.2341853371067\\
10.4	5.98189257540055\\
11.6	5.98189257540055\\
11.6	4.2341853371067\\
10.4	4.2341853371067\\
};
\addplot [color=clr4, forget plot]
  table[row sep=crcr]{%
12.8	2.92640747675317\\
12.8	3.38597023786733\\
14	3.38597023786733\\
14	2.92640747675317\\
12.8	2.92640747675317\\
};
\addplot [color=clr3, forget plot]
  table[row sep=crcr]{%
15.2	2.32858418595226\\
15.2	2.74754824964853\\
16.4	2.74754824964853\\
16.4	2.32858418595226\\
15.2	2.32858418595226\\
};
\addplot [color=clr9, forget plot]
  table[row sep=crcr]{%
20.4	4.3949120603055\\
20.4	6.15452938201017\\
21.6	6.15452938201017\\
21.6	4.3949120603055\\
20.4	4.3949120603055\\
};
\addplot [color=clr4, forget plot]
  table[row sep=crcr]{%
22.8	3.03980573071465\\
22.8	3.52923784621939\\
24	3.52923784621939\\
24	3.03980573071465\\
22.8	3.03980573071465\\
};
\addplot [color=clr3, forget plot]
  table[row sep=crcr]{%
25.2	2.42224997460902\\
25.2	2.87742626213708\\
26.4	2.87742626213708\\
26.4	2.42224997460902\\
25.2	2.42224997460902\\
};
\addplot [color=clr9, forget plot]
  table[row sep=crcr]{%
30.4	4.40555252234057\\
30.4	6.16895525981445\\
31.6	6.16895525981445\\
31.6	4.40555252234057\\
30.4	4.40555252234057\\
};
\addplot [color=clr4, forget plot]
  table[row sep=crcr]{%
32.8	3.06773317688859\\
32.8	3.55552571978686\\
34	3.55552571978686\\
34	3.06773317688859\\
32.8	3.06773317688859\\
};
\addplot [color=clr3, forget plot]
  table[row sep=crcr]{%
35.2	2.49199858010791\\
35.2	2.94084294219502\\
36.4	2.94084294219502\\
36.4	2.49199858010791\\
35.2	2.49199858010791\\
};
\addplot [color=clr9, forget plot]
  table[row sep=crcr]{%
40.4	4.53481749169849\\
40.4	6.35518602965177\\
41.6	6.35518602965177\\
41.6	4.53481749169849\\
40.4	4.53481749169849\\
};
\addplot [color=clr4, forget plot]
  table[row sep=crcr]{%
42.8	3.09155458041362\\
42.8	3.56540933746392\\
44	3.56540933746392\\
44	3.09155458041362\\
42.8	3.09155458041362\\
};
\addplot [color=clr3, forget plot]
  table[row sep=crcr]{%
45.2	2.58616117298037\\
45.2	3.03555441226688\\
46.4	3.03555441226688\\
46.4	2.58616117298037\\
45.2	2.58616117298037\\
};
\addplot [color=clr9, forget plot]
  table[row sep=crcr]{%
0.4	5.16293541658298\\
1.6	5.16293541658298\\
};
\addplot [color=clr4, forget plot]
  table[row sep=crcr]{%
2.8	2.56340389664959\\
4	2.56340389664959\\
};
\addplot [color=clr3, forget plot]
  table[row sep=crcr]{%
5.2	2.4603717303852\\
6.4	2.4603717303852\\
};
\addplot [color=clr9, forget plot]
  table[row sep=crcr]{%
10.4	5.18822993115038\\
11.6	5.18822993115038\\
};
\addplot [color=clr4, forget plot]
  table[row sep=crcr]{%
12.8	3.15881998110338\\
14	3.15881998110338\\
};
\addplot [color=clr3, forget plot]
  table[row sep=crcr]{%
15.2	2.50997960849938\\
16.4	2.50997960849938\\
};
\addplot [color=clr9, forget plot]
  table[row sep=crcr]{%
20.4	5.35686966911181\\
21.6	5.35686966911181\\
};
\addplot [color=clr4, forget plot]
  table[row sep=crcr]{%
22.8	3.26597208946539\\
24	3.26597208946539\\
};
\addplot [color=clr3, forget plot]
  table[row sep=crcr]{%
25.2	2.63600916110704\\
26.4	2.63600916110704\\
};
\addplot [color=clr9, forget plot]
  table[row sep=crcr]{%
30.4	5.42975533849259\\
31.6	5.42975533849259\\
};
\addplot [color=clr4, forget plot]
  table[row sep=crcr]{%
32.8	3.29791727440407\\
34	3.29791727440407\\
};
\addplot [color=clr3, forget plot]
  table[row sep=crcr]{%
35.2	2.70081408933638\\
36.4	2.70081408933638\\
};
\addplot [color=clr9]
  table[row sep=crcr]{%
40.4	5.49003487604118\\
41.6	5.49003487604118\\
};
\addlegendentry{Eq}

\addplot [color=clr4]
  table[row sep=crcr]{%
42.8	3.29796384976617\\
44	3.29796384976617\\
};
\addlegendentry{Rb}

\addplot [color=clr3]
  table[row sep=crcr]{%
45.2	2.77821450491231\\
46.4	2.77821450491231\\
};
\addlegendentry{Opt}

\addplot [color=black, draw=none, mark=+, mark options={solid, red}, forget plot]
  table[row sep=crcr]{%
nan	nan\\
};
\addplot [color=black, draw=none, mark=+, mark options={solid, red}, forget plot]
  table[row sep=crcr]{%
3.4	3.51173503461457\\
3.4	3.52469119100652\\
3.4	3.52549292145308\\
3.4	3.53085509320432\\
3.4	3.53933506181516\\
3.4	3.54835243477331\\
3.4	3.55692805884791\\
3.4	3.64818315106197\\
3.4	3.65093494467303\\
3.4	3.66318546695445\\
3.4	3.66361097426672\\
3.4	3.67166702919825\\
3.4	3.68863713077963\\
3.4	3.68948616099467\\
3.4	3.69655121632228\\
3.4	3.69719555530912\\
3.4	3.71212535682176\\
3.4	3.71570811888512\\
3.4	3.71714617235413\\
3.4	3.72363516088031\\
3.4	3.77471950136798\\
3.4	3.81219522580113\\
3.4	3.85863929023205\\
3.4	3.86785259026669\\
3.4	3.88750345209641\\
3.4	3.93588363390714\\
3.4	3.93651284820369\\
3.4	3.94146355294185\\
};
\addplot [color=black, draw=none, mark=+, mark options={solid, red}, forget plot]
  table[row sep=crcr]{%
5.8	3.31271089715238\\
5.8	3.31615785923506\\
5.8	3.34207905567245\\
5.8	3.35262867180546\\
5.8	3.36495994195544\\
5.8	3.37115139986601\\
5.8	3.38413726354384\\
5.8	3.40040866088101\\
5.8	3.44408032743383\\
5.8	3.44616190178701\\
5.8	3.45891874087086\\
5.8	3.46474988894988\\
5.8	3.47172794536056\\
5.8	3.47202207145569\\
5.8	3.47537585406677\\
5.8	3.4767081852448\\
5.8	3.47673398845775\\
5.8	3.51806980284905\\
5.8	3.54021011267185\\
5.8	3.55615812353434\\
5.8	3.5709015169299\\
5.8	3.57185235272133\\
5.8	3.57270110394119\\
5.8	3.59469033157476\\
5.8	3.59762794031647\\
5.8	3.64714896582704\\
5.8	3.67398809398593\\
5.8	3.6870804694866\\
5.8	3.69895815437661\\
5.8	3.72338076197495\\
5.8	3.78281983015416\\
5.8	3.80376595331632\\
};
\addplot [color=black, draw=none, mark=+, mark options={solid, red}, forget plot]
  table[row sep=crcr]{%
nan	nan\\
};
\addplot [color=black, draw=none, mark=+, mark options={solid, red}, forget plot]
  table[row sep=crcr]{%
13.4	4.07912504823054\\
13.4	4.08146044306775\\
13.4	4.09326922964237\\
13.4	4.11609623751803\\
13.4	4.13058109461006\\
13.4	4.14690425467847\\
13.4	4.1776124874316\\
13.4	4.2146965413011\\
13.4	4.24554823607678\\
13.4	4.29531033045215\\
13.4	4.30500681526161\\
13.4	4.3103149562371\\
13.4	4.32056712475049\\
13.4	4.34775361710472\\
13.4	4.37419998482099\\
13.4	4.41580319663661\\
13.4	4.4223110986009\\
13.4	4.4862140691241\\
13.4	4.49636862421256\\
13.4	4.51957680174457\\
13.4	4.57315776458649\\
13.4	4.60379263338369\\
};
\addplot [color=black, draw=none, mark=+, mark options={solid, red}, forget plot]
  table[row sep=crcr]{%
15.8	3.38585725632692\\
15.8	3.39284407964658\\
15.8	3.3959992569413\\
15.8	3.42537642083646\\
15.8	3.43425435444445\\
15.8	3.46502658559717\\
15.8	3.49895502547296\\
15.8	3.52336559664831\\
15.8	3.56115604900833\\
15.8	3.58532451275505\\
15.8	3.59365096063818\\
15.8	3.60370733711145\\
15.8	3.61238105203286\\
15.8	3.61360121284086\\
15.8	3.64905077243199\\
15.8	3.6760737988238\\
15.8	3.69513410534053\\
15.8	3.84959130232177\\
15.8	3.91411378818213\\
15.8	3.98514408834145\\
15.8	3.99300496582144\\
15.8	4.01064876223253\\
};
\addplot [color=black, draw=none, mark=+, mark options={solid, red}, forget plot]
  table[row sep=crcr]{%
nan	nan\\
};
\addplot [color=black, draw=none, mark=+, mark options={solid, red}, forget plot]
  table[row sep=crcr]{%
23.4	4.28076184420902\\
23.4	4.28481196806792\\
23.4	4.36844848687602\\
23.4	4.37122088029839\\
23.4	4.38016487491216\\
23.4	4.54665308524747\\
23.4	4.57397869164558\\
23.4	4.77331849306552\\
};
\addplot [color=black, draw=none, mark=+, mark options={solid, red}, forget plot]
  table[row sep=crcr]{%
25.8	3.57948513710816\\
25.8	3.59496311501391\\
25.8	3.61957645419841\\
25.8	3.65881530889082\\
25.8	3.66371541665248\\
25.8	3.66602898496292\\
25.8	3.68955199803261\\
25.8	3.6979392791294\\
25.8	3.72996607904492\\
25.8	3.73630550139553\\
25.8	3.73703987963844\\
25.8	3.88565873633461\\
25.8	4.08425217873959\\
};
\addplot [color=black, draw=none, mark=+, mark options={solid, red}, forget plot]
  table[row sep=crcr]{%
nan	nan\\
};
\addplot [color=black, draw=none, mark=+, mark options={solid, red}, forget plot]
  table[row sep=crcr]{%
33.4	4.30422647045433\\
33.4	4.32737046971763\\
33.4	4.33353889211839\\
33.4	4.3346224385037\\
33.4	4.35749133257848\\
33.4	4.45216780978927\\
33.4	4.69745327829307\\
33.4	4.8931540321145\\
33.4	5.14058130649101\\
33.4	5.28789772004812\\
33.4	5.42169527254271\\
};
\addplot [color=black, draw=none, mark=+, mark options={solid, red}, forget plot]
  table[row sep=crcr]{%
35.8	3.62324156560605\\
35.8	3.6255922087108\\
35.8	3.63784545021308\\
35.8	3.64600180657517\\
35.8	3.65377579233597\\
35.8	3.65898225398282\\
35.8	3.65911219357691\\
35.8	3.66800553778351\\
35.8	3.67702858652172\\
35.8	3.68669742844911\\
35.8	3.69107725836386\\
35.8	3.74323782672501\\
35.8	3.7761021786941\\
35.8	3.79510224973684\\
35.8	3.80007526890697\\
35.8	3.85663391889869\\
35.8	4.0377543258535\\
35.8	4.09212394571432\\
35.8	4.51886804673453\\
35.8	4.70252322170636\\
35.8	4.7183044303366\\
};
\addplot [color=black, draw=none, mark=+, mark options={solid, red}, forget plot]
  table[row sep=crcr]{%
nan	nan\\
};
\addplot [color=black, draw=none, mark=+, mark options={solid, red}, forget plot]
  table[row sep=crcr]{%
43.4	4.28153610929718\\
43.4	4.31692026630635\\
43.4	4.34224436811101\\
43.4	4.3592277590406\\
43.4	4.36311297946493\\
43.4	4.37460605714019\\
43.4	4.43308756877968\\
43.4	4.44625768980956\\
43.4	4.46118487428104\\
43.4	4.46222524714221\\
43.4	4.48679529146302\\
43.4	4.52489747011216\\
43.4	4.74007567037052\\
};
\addplot [color=black, draw=none, mark=+, mark options={solid, red}, forget plot]
  table[row sep=crcr]{%
45.8	3.71089912697244\\
45.8	3.71129949618176\\
45.8	3.71713056217562\\
45.8	3.72979997204846\\
45.8	3.73685106970093\\
45.8	3.76759701868328\\
45.8	3.77308541374133\\
45.8	3.78891216881747\\
45.8	3.79554686268411\\
45.8	3.82408869968113\\
45.8	3.85640159655699\\
45.8	3.87939417255824\\
45.8	3.89549933376863\\
45.8	3.92325393688343\\
45.8	3.93040652424857\\
45.8	3.94361856112978\\
45.8	4.09292064700855\\
};
\end{axis}
\end{tikzpicture}%

%% file: Figures/load_boxplot.tex
\begin{tikzpicture} 
\input{doc/colors.tex}
font = \footnotesize, 
\begin{axis}[legend pos=south west, scale only axis, 
yticklabel style={
      /pgf/number format/fixed,
      /pgf/number format/precision=5
},
scaled y ticks=false, 
width=0.39\columnwidth,
height=0.3\columnwidth,
scale only axis,
unbounded coords=jump,
xmin=-0.199999999999999,
xmax=47,
xtick={3.5,13.5,23.5,33.5,43.5},
xticklabels={{1},{2},{3},{4},{5}},
xlabel style={font=\color{white!15!black}},
xlabel={Number of nodes},
ymin=1,
ymax=9.14040893512376,
ylabel style={font=\color{white!15!black}},
ylabel={Download time (s)},
compat=1.3,
legend style={/tikz/every even column/.append style={column sep=0.2cm}},
legend style={{anchor=west},at={(0.5, 1.15)}, anchor=north, fill=white, draw=none, legend columns=-1, font = \footnotesize},
every axis plot/.append style={fill,fill opacity=0.2}
]

\addplot [color= clr9,   forget plot]
  table[row sep=crcr]{%
1	5.82715531747374\\
1	6.60934608183295\\
};
\addplot [color= clr4,   forget plot]
  table[row sep=crcr]{%
3.4	2.8182663005433\\
3.4	3.50416417951482\\
};
\addplot [color= clr3,   forget plot]
  table[row sep=crcr]{%
5.8	2.67858599772455\\
5.8	3.29305488297482\\
};
\addplot [color= clr9,   forget plot]
  table[row sep=crcr]{%
11	6.03124767795987\\
11	6.60868877955217\\
};
\addplot [color= clr4,   forget plot]
  table[row sep=crcr]{%
13.4	2.82785580616443\\
13.4	3.525651427601\\
};
\addplot [color= clr3,   forget plot]
  table[row sep=crcr]{%
15.8	2.69203397026607\\
15.8	3.35103752442224\\
};
\addplot [color= clr9,   forget plot]
  table[row sep=crcr]{%
21	6.08581061612562\\
21	7.4728997777728\\
};
\addplot [color= clr4,   forget plot]
  table[row sep=crcr]{%
23.4	3.11367003728995\\
23.4	3.91296731161431\\
};
\addplot [color= clr3,   forget plot]
  table[row sep=crcr]{%
25.8	2.71583955687121\\
25.8	3.33846334528758\\
};
\addplot [color= clr9,   forget plot]
  table[row sep=crcr]{%
31	6.06493689393145\\
31	7.33488222749178\\
};
\addplot [color= clr4,   forget plot]
  table[row sep=crcr]{%
33.4	3.20655750509822\\
33.4	3.96698181129911\\
};
\addplot [color= clr3,   forget plot]
  table[row sep=crcr]{%
35.8	2.7546637870977\\
35.8	3.40735711414922\\
};
\addplot [color= clr9,   forget plot]
  table[row sep=crcr]{%
41	6.16097352306099\\
41	8.46602458918355\\
};
\addplot [color= clr4,   forget plot]
  table[row sep=crcr]{%
43.4	3.85515695460873\\
43.4	5.21131643521085\\
};
\addplot [color= clr3,   forget plot]
  table[row sep=crcr]{%
45.8	2.88736648785724\\
45.8	3.56475286637753\\
};
\addplot [color= clr9,   forget plot]
  table[row sep=crcr]{%
1	2.21106013712771\\
1	4.17948315148649\\
};
\addplot [color= clr4,   forget plot]
  table[row sep=crcr]{%
3.4	1.85890665410916\\
3.4	2.34661879858439\\
};
\addplot [color= clr3,   forget plot]
  table[row sep=crcr]{%
5.8	1.77890001073338\\
5.8	2.25731589828749\\
};
\addplot [color= clr9,   forget plot]
  table[row sep=crcr]{%
11	2.06074512441265\\
11	4.18357650067212\\
};
\addplot [color= clr4,   forget plot]
  table[row sep=crcr]{%
13.4	1.8815470566752\\
13.4	2.35088686576319\\
};
\addplot [color= clr3,   forget plot]
  table[row sep=crcr]{%
15.8	1.77731591758379\\
15.8	2.25255545646641\\
};
\addplot [color= clr9,   forget plot]
  table[row sep=crcr]{%
21	1.90132314066204\\
21	4.29068544328613\\
};
\addplot [color= clr4,   forget plot]
  table[row sep=crcr]{%
23.4	2.02338952078794\\
23.4	2.57568825924727\\
};
\addplot [color= clr3,   forget plot]
  table[row sep=crcr]{%
25.8	1.72577396508581\\
25.8	2.2674749672752\\
};
\addplot [color= clr9,   forget plot]
  table[row sep=crcr]{%
31	2.09283446581448\\
31	4.21274074974015\\
};
\addplot [color= clr4,   forget plot]
  table[row sep=crcr]{%
33.4	2.07888578229604\\
33.4	2.6891950676009\\
};
\addplot [color= clr3,   forget plot]
  table[row sep=crcr]{%
35.8	1.7854572149382\\
35.8	2.31671010200834\\
};
\addplot [color= clr9,   forget plot]
  table[row sep=crcr]{%
41	2.03721749664205\\
41	4.42745831838501\\
};
\addplot [color= clr4,   forget plot]
  table[row sep=crcr]{%
43.4	2.27222082725317\\
43.4	2.93936482009493\\
};
\addplot [color= clr3,   forget plot]
  table[row sep=crcr]{%
45.8	1.85971112785636\\
45.8	2.43143029162424\\
};
\addplot [color= clr9, forget plot]
  table[row sep=crcr]{%
0.7	6.60934608183295\\
1.3	6.60934608183295\\
};
\addplot [color= clr4, forget plot]
  table[row sep=crcr]{%
3.1	3.50416417951482\\
3.7	3.50416417951482\\
};
\addplot [color= clr3, forget plot]
  table[row sep=crcr]{%
5.5	3.29305488297482\\
6.1	3.29305488297482\\
};
\addplot [color= clr9, forget plot]
  table[row sep=crcr]{%
10.7	6.60868877955217\\
11.3	6.60868877955217\\
};
\addplot [color= clr4, forget plot]
  table[row sep=crcr]{%
13.1	3.525651427601\\
13.7	3.525651427601\\
};
\addplot [color= clr3, forget plot]
  table[row sep=crcr]{%
15.5	3.35103752442224\\
16.1	3.35103752442224\\
};
\addplot [color= clr9, forget plot]
  table[row sep=crcr]{%
20.7	7.4728997777728\\
21.3	7.4728997777728\\
};
\addplot [color= clr4, forget plot]
  table[row sep=crcr]{%
23.1	3.91296731161431\\
23.7	3.91296731161431\\
};
\addplot [color= clr3, forget plot]
  table[row sep=crcr]{%
25.5	3.33846334528758\\
26.1	3.33846334528758\\
};
\addplot [color= clr9, forget plot]
  table[row sep=crcr]{%
30.7	7.33488222749178\\
31.3	7.33488222749178\\
};
\addplot [color= clr4, forget plot]
  table[row sep=crcr]{%
33.1	3.96698181129911\\
33.7	3.96698181129911\\
};
\addplot [color= clr3, forget plot]
  table[row sep=crcr]{%
35.5	3.40735711414922\\
36.1	3.40735711414922\\
};
\addplot [color= clr9, forget plot]
  table[row sep=crcr]{%
40.7	8.46602458918355\\
41.3	8.46602458918355\\
};
\addplot [color= clr4, forget plot]
  table[row sep=crcr]{%
43.1	5.21131643521085\\
43.7	5.21131643521085\\
};
\addplot [color= clr3, forget plot]
  table[row sep=crcr]{%
45.5	3.56475286637753\\
46.1	3.56475286637753\\
};
\addplot [color= clr9, forget plot]
  table[row sep=crcr]{%
0.7	2.21106013712771\\
1.3	2.21106013712771\\
};
\addplot [color= clr4, forget plot]
  table[row sep=crcr]{%
3.1	1.85890665410916\\
3.7	1.85890665410916\\
};
\addplot [color= clr3, forget plot]
  table[row sep=crcr]{%
5.5	1.77890001073338\\
6.1	1.77890001073338\\
};
\addplot [color= clr9, forget plot]
  table[row sep=crcr]{%
10.7	2.06074512441265\\
11.3	2.06074512441265\\
};
\addplot [color= clr4, forget plot]
  table[row sep=crcr]{%
13.1	1.8815470566752\\
13.7	1.8815470566752\\
};
\addplot [color= clr3, forget plot]
  table[row sep=crcr]{%
15.5	1.77731591758379\\
16.1	1.77731591758379\\
};
\addplot [color= clr9, forget plot]
  table[row sep=crcr]{%
20.7	1.90132314066204\\
21.3	1.90132314066204\\
};
\addplot [color= clr4, forget plot]
  table[row sep=crcr]{%
23.1	2.02338952078794\\
23.7	2.02338952078794\\
};
\addplot [color= clr3, forget plot]
  table[row sep=crcr]{%
25.5	1.72577396508581\\
26.1	1.72577396508581\\
};
\addplot [color= clr9, forget plot]
  table[row sep=crcr]{%
30.7	2.09283446581448\\
31.3	2.09283446581448\\
};
\addplot [color= clr4, forget plot]
  table[row sep=crcr]{%
33.1	2.07888578229604\\
33.7	2.07888578229604\\
};
\addplot [color= clr3, forget plot]
  table[row sep=crcr]{%
35.5	1.7854572149382\\
36.1	1.7854572149382\\
};
\addplot [color= clr9, forget plot]
  table[row sep=crcr]{%
40.7	2.03721749664205\\
41.3	2.03721749664205\\
};
\addplot [color= clr4, forget plot]
  table[row sep=crcr]{%
43.1	2.27222082725317\\
43.7	2.27222082725317\\
};
\addplot [color= clr3, forget plot]
  table[row sep=crcr]{%
45.5	1.85971112785636\\
46.1	1.85971112785636\\
};
\addplot [color= clr9, forget plot]
  table[row sep=crcr]{%
0.4	4.17948315148649\\
0.4	5.82715531747374\\
1.6	5.82715531747374\\
1.6	4.17948315148649\\
0.4	4.17948315148649\\
};
\addplot [color= clr4, forget plot]
  table[row sep=crcr]{%
2.8	2.34661879858439\\
2.8	2.8182663005433\\
4	2.8182663005433\\
4	2.34661879858439\\
2.8	2.34661879858439\\
};
\addplot [color= clr3, forget plot]
  table[row sep=crcr]{%
5.2	2.25731589828749\\
5.2	2.67858599772455\\
6.4	2.67858599772455\\
6.4	2.25731589828749\\
5.2	2.25731589828749\\
};
\addplot [color= clr9, forget plot]
  table[row sep=crcr]{%
10.4	4.18357650067212\\
10.4	6.03124767795987\\
11.6	6.03124767795987\\
11.6	4.18357650067212\\
10.4	4.18357650067212\\
};
\addplot [color= clr4, forget plot]
  table[row sep=crcr]{%
12.8	2.35088686576319\\
12.8	2.82785580616443\\
14	2.82785580616443\\
14	2.35088686576319\\
12.8	2.35088686576319\\
};
\addplot [color= clr3, forget plot]
  table[row sep=crcr]{%
15.2	2.25255545646641\\
15.2	2.69203397026607\\
16.4	2.69203397026607\\
16.4	2.25255545646641\\
15.2	2.25255545646641\\
};
\addplot [color= clr9, forget plot]
  table[row sep=crcr]{%
20.4	4.29068544328613\\
20.4	6.08581061612562\\
21.6	6.08581061612562\\
21.6	4.29068544328613\\
20.4	4.29068544328613\\
};
\addplot [color= clr4, forget plot]
  table[row sep=crcr]{%
22.8	2.57568825924727\\
22.8	3.11367003728995\\
24	3.11367003728995\\
24	2.57568825924727\\
22.8	2.57568825924727\\
};
\addplot [color= clr3, forget plot]
  table[row sep=crcr]{%
25.2	2.2674749672752\\
25.2	2.71583955687121\\
26.4	2.71583955687121\\
26.4	2.2674749672752\\
25.2	2.2674749672752\\
};
\addplot [color= clr9, forget plot]
  table[row sep=crcr]{%
30.4	4.21274074974015\\
30.4	6.06493689393145\\
31.6	6.06493689393145\\
31.6	4.21274074974015\\
30.4	4.21274074974015\\
};
\addplot [color= clr4, forget plot]
  table[row sep=crcr]{%
32.8	2.6891950676009\\
32.8	3.20655750509822\\
34	3.20655750509822\\
34	2.6891950676009\\
32.8	2.6891950676009\\
};
\addplot [color= clr3, forget plot]
  table[row sep=crcr]{%
35.2	2.31671010200834\\
35.2	2.7546637870977\\
36.4	2.7546637870977\\
36.4	2.31671010200834\\
35.2	2.31671010200834\\
};
\addplot [color= clr9, forget plot]
  table[row sep=crcr]{%
40.4	4.42745831838501\\
40.4	6.16097352306099\\
41.6	6.16097352306099\\
41.6	4.42745831838501\\
40.4	4.42745831838501\\
};
\addplot [color= clr4, forget plot]
  table[row sep=crcr]{%
42.8	2.93936482009493\\
42.8	3.85515695460873\\
44	3.85515695460873\\
44	2.93936482009493\\
42.8	2.93936482009493\\
};
\addplot [color= clr3, forget plot]
  table[row sep=crcr]{%
45.2	2.43143029162424\\
45.2	2.88736648785724\\
46.4	2.88736648785724\\
46.4	2.43143029162424\\
45.2	2.43143029162424\\
};
\addplot [color= clr9, forget plot]
  table[row sep=crcr]{%
0.4	5.17038097995495\\
1.6	5.17038097995495\\
};
\addplot [color= clr4, forget plot]
  table[row sep=crcr]{%
2.8	2.54975777987275\\
4	2.54975777987275\\
};
\addplot [color= clr3, forget plot]
  table[row sep=crcr]{%
5.2	2.43712279545954\\
6.4	2.43712279545954\\
};
\addplot [color= clr9, forget plot]
  table[row sep=crcr]{%
10.4	5.18616960644649\\
11.6	5.18616960644649\\
};
\addplot [color= clr4, forget plot]
  table[row sep=crcr]{%
12.8	2.56226736076585\\
14	2.56226736076585\\
};
\addplot [color= clr3, forget plot]
  table[row sep=crcr]{%
15.2	2.45113831999067\\
16.4	2.45113831999067\\
};
\addplot [color= clr9, forget plot]
  table[row sep=crcr]{%
20.4	5.19243087053393\\
21.6	5.19243087053393\\
};
\addplot [color= clr4, forget plot]
  table[row sep=crcr]{%
22.8	2.82710413427399\\
24	2.82710413427399\\
};
\addplot [color= clr3, forget plot]
  table[row sep=crcr]{%
25.2	2.47387362487472\\
26.4	2.47387362487472\\
};
\addplot [color= clr9, forget plot]
  table[row sep=crcr]{%
30.4	5.13228526873838\\
31.6	5.13228526873838\\
};
\addplot [color= clr4, forget plot]
  table[row sep=crcr]{%
32.8	2.94984994819753\\
34	2.94984994819753\\
};
\addplot [color= clr3, forget plot]
  table[row sep=crcr]{%
35.2	2.51732168184965\\
36.4	2.51732168184965\\
};
\addplot [color= clr9]
  table[row sep=crcr]{%
40.4	5.36452359128844\\
41.6	5.36452359128844\\
};
\addlegendentry{Eq}

\addplot [color= clr4]
  table[row sep=crcr]{%
42.8	3.36438234587796\\
44	3.36438234587796\\
};
\addlegendentry{Rb}

\addplot [color= clr3]
  table[row sep=crcr]{%
45.2	2.63691190369878\\
46.4	2.63691190369878\\
};
\addlegendentry{Opt}

\addplot [color=black, draw=none, mark=+, mark options={solid, red}, forget plot]
  table[row sep=crcr]{%
nan	nan\\
};
\addplot [color=black, draw=none, mark=+, mark options={solid, red}, forget plot]
  table[row sep=crcr]{%
3.4	3.53031782701322\\
3.4	3.53125153399303\\
3.4	3.53239721704859\\
3.4	3.56636097390514\\
3.4	3.60941342230924\\
3.4	3.61019382365885\\
3.4	3.63086803913705\\
3.4	3.64977091038592\\
3.4	3.65734684330829\\
3.4	3.66690189005144\\
3.4	3.68232689117435\\
3.4	3.78769936921322\\
3.4	3.84493124822184\\
3.4	3.84641315772113\\
3.4	3.88431543865477\\
3.4	3.91370345528511\\
3.4	4.05833997323853\\
3.4	4.15397324973852\\
};
\addplot [color=black, draw=none, mark=+, mark options={solid, red}, forget plot]
  table[row sep=crcr]{%
5.8	3.32325993738326\\
5.8	3.33059144773199\\
5.8	3.33332587727296\\
5.8	3.3379151069301\\
5.8	3.34063094126203\\
5.8	3.34081732070289\\
5.8	3.34930706830972\\
5.8	3.36004006158886\\
5.8	3.36498586626592\\
5.8	3.38282424742123\\
5.8	3.38577986380809\\
5.8	3.40264870291282\\
5.8	3.40667877902379\\
5.8	3.4467387817805\\
5.8	3.44702315588391\\
5.8	3.44770080693054\\
5.8	3.45701471820042\\
5.8	3.4586246107284\\
5.8	3.46553608871505\\
5.8	3.53751877352912\\
5.8	3.55034736006595\\
5.8	3.62102235101754\\
5.8	3.62544050302738\\
5.8	3.65434299913503\\
5.8	3.72753916367599\\
5.8	3.79498124739693\\
5.8	3.95815072624035\\
5.8	3.99464340893254\\
};
\addplot [color=black, draw=none, mark=+, mark options={solid, red}, forget plot]
  table[row sep=crcr]{%
nan	nan\\
};
\addplot [color=black, draw=none, mark=+, mark options={solid, red}, forget plot]
  table[row sep=crcr]{%
13.4	3.55655022726312\\
13.4	3.55846795037018\\
13.4	3.56057796838672\\
13.4	3.59678528477881\\
13.4	3.59959102001053\\
13.4	3.61056584325813\\
13.4	3.65904030224047\\
13.4	3.69708112400325\\
13.4	3.75318184332026\\
13.4	3.75620488736255\\
13.4	3.84468092410736\\
13.4	3.89880842008711\\
13.4	3.97121224506208\\
};
\addplot [color=black, draw=none, mark=+, mark options={solid, red}, forget plot]
  table[row sep=crcr]{%
15.8	3.36450379068403\\
15.8	3.37343784586973\\
15.8	3.37588043588265\\
15.8	3.40381474843468\\
15.8	3.4182795268442\\
15.8	3.44239110179254\\
15.8	3.48513400534144\\
15.8	3.49995259949468\\
15.8	3.50344878321472\\
15.8	3.50371599365044\\
15.8	3.52102459141099\\
15.8	3.55146320633175\\
15.8	3.56033227867335\\
15.8	3.63090245232919\\
15.8	3.72181549152202\\
15.8	3.76540856798178\\
15.8	3.87332487433058\\
};
\addplot [color=black, draw=none, mark=+, mark options={solid, red}, forget plot]
  table[row sep=crcr]{%
nan	nan\\
};
\addplot [color=black, draw=none, mark=+, mark options={solid, red}, forget plot]
  table[row sep=crcr]{%
23.4	3.9210454180741\\
23.4	3.92709785842409\\
23.4	3.94016348301112\\
23.4	3.9446362757032\\
23.4	3.96930433816044\\
23.4	3.99877212388573\\
23.4	4.06408916410695\\
23.4	4.06850898845143\\
23.4	4.08562289009043\\
23.4	4.10673591961411\\
23.4	4.14314547470213\\
23.4	4.19650014837188\\
23.4	4.4260904662965\\
};
\addplot [color=black, draw=none, mark=+, mark options={solid, red}, forget plot]
  table[row sep=crcr]{%
25.8	3.39524191655855\\
25.8	3.40421560618084\\
25.8	3.40652100819097\\
25.8	3.40878116877413\\
25.8	3.42526933316748\\
25.8	3.49236153627901\\
25.8	3.50340780101111\\
25.8	3.60604871996824\\
25.8	3.6098353954828\\
25.8	3.70092356451289\\
25.8	3.75367837638671\\
25.8	3.76187503609957\\
25.8	3.81783367414909\\
};
\addplot [color=black, draw=none, mark=+, mark options={solid, red}, forget plot]
  table[row sep=crcr]{%
nan	nan\\
};
\addplot [color=black, draw=none, mark=+, mark options={solid, red}, forget plot]
  table[row sep=crcr]{%
33.4	3.9941262600379\\
33.4	4.00491961095817\\
33.4	4.04054081620154\\
33.4	4.10332911171881\\
33.4	4.1101870223521\\
33.4	4.15864610518115\\
33.4	4.17091051265332\\
33.4	4.20267699220965\\
33.4	4.25681625080433\\
33.4	4.26904355433852\\
33.4	4.42503535728324\\
};
\addplot [color=black, draw=none, mark=+, mark options={solid, red}, forget plot]
  table[row sep=crcr]{%
35.8	3.4278450007367\\
35.8	3.42906994699282\\
35.8	3.44855951711821\\
35.8	3.46699945704845\\
35.8	3.47457367994152\\
35.8	3.48003525567589\\
35.8	3.49225473829469\\
35.8	3.51942951248267\\
35.8	3.56165480018164\\
35.8	3.59430549467617\\
35.8	3.63479497777164\\
35.8	3.69145094986309\\
35.8	3.7198636628161\\
35.8	3.72107080315312\\
35.8	3.74049225901778\\
35.8	3.74897627096122\\
35.8	3.77230239821431\\
};
\addplot [color=black, draw=none, mark=+, mark options={solid, red}, forget plot]
  table[row sep=crcr]{%
41	8.78733107940767\\
};
\addplot [color=black, draw=none, mark=+, mark options={solid, red}, forget plot]
  table[row sep=crcr]{%
43.4	5.23229052934373\\
43.4	5.24017670995045\\
43.4	5.24685088746786\\
43.4	5.25159994570391\\
43.4	5.29335859453751\\
43.4	5.32448925527838\\
43.4	5.41174479532487\\
43.4	5.42397077303374\\
43.4	5.79026584796559\\
43.4	5.84504519438004\\
43.4	6.120664412741\\
};
\addplot [color=black, draw=none, mark=+, mark options={solid, red}, forget plot]
  table[row sep=crcr]{%
45.8	3.60005091993899\\
45.8	3.60448034597283\\
45.8	3.62082714849155\\
45.8	3.63801662734252\\
45.8	3.63981198078551\\
45.8	3.65214137562829\\
45.8	3.65958001128362\\
45.8	3.66245421875495\\
45.8	3.67770171513275\\
45.8	3.68253832160058\\
45.8	3.70514760777851\\
45.8	3.7085979452608\\
45.8	3.71587664208433\\
45.8	3.79670511595469\\
45.8	3.80412409518296\\
45.8	3.81372537120858\\
45.8	3.82081964841459\\
45.8	3.83405615310563\\
45.8	3.85642787181389\\
45.8	4.00828222742551\\
45.8	4.07214410910427\\
45.8	4.10318385834945\\
45.8	4.13072861552648\\
45.8	4.327158528425\\
};
\end{axis}
\end{tikzpicture}%

%% file: Figures/overload_boxplot.tex
\begin{tikzpicture} 
\input{doc/colors.tex}
font = \footnotesize, 
\begin{axis}[legend pos=south west, scale only axis, 
yticklabel style={
      /pgf/number format/fixed,
      /pgf/number format/precision=5
},
scaled y ticks=false, 
width=0.39\columnwidth,
height=0.3\columnwidth,
scale only axis,
unbounded coords=jump,
xmin=-0.199999999999999,
xmax=47,
xtick={3.5,13.5,23.5,33.5,43.5},
xticklabels={{1},{2},{3},{4},{5}},
xlabel style={font=\color{white!15!black}},
xlabel={Number of nodes},
ymin=1,
ymax=12,
ytick={3,6,9,12},
yticklabels={3,6,9,12},
ylabel style={font=\color{white!15!black}},
ylabel={Download time (s)},
compat=1.3,
legend style={/tikz/every even column/.append style={column sep=0.2cm}},
legend style={{anchor=west},at={(0.5, 1.15)}, anchor=north, fill=white, draw=none, legend columns=-1, font = \footnotesize},
every axis plot/.append style={fill,fill opacity=0.2}
]

\addplot [color=clr9,   forget plot]
  table[row sep=crcr]{%
1	6.12968998427154\\
1	6.61642465437895\\
};
\addplot [color=clr4,   forget plot]
  table[row sep=crcr]{%
3.4	2.80943820974808\\
3.4	3.48522957647521\\
};
\addplot [color=clr3,   forget plot]
  table[row sep=crcr]{%
5.8	2.66776191810686\\
5.8	3.29909771967089\\
};
\addplot [color=clr9,   forget plot]
  table[row sep=crcr]{%
11	6.04127156228546\\
11	6.61158701824564\\
};
\addplot [color=clr4,   forget plot]
  table[row sep=crcr]{%
13.4	2.82053930391937\\
13.4	3.55036574457205\\
};
\addplot [color=clr3,   forget plot]
  table[row sep=crcr]{%
15.8	2.68861593309725\\
15.8	3.37606640715711\\
};
\addplot [color=clr9,   forget plot]
  table[row sep=crcr]{%
21	8.35143424003738\\
21	10.7416425125128\\
};
\addplot [color=clr4,   forget plot]
  table[row sep=crcr]{%
23.4	7.60819168686114\\
23.4	8.29058053647724\\
};
\addplot [color=clr3,   forget plot]
  table[row sep=crcr]{%
25.8	5.08099999870758\\
25.8	5.09999999999146\\
};
\addplot [color=clr9,   forget plot]
  table[row sep=crcr]{%
31	9.33318724276664\\
31	11.4982329188214\\
};
\addplot [color=clr4,   forget plot]
  table[row sep=crcr]{%
33.4	7.63775210788479\\
33.4	8.29823291882136\\
};
\addplot [color=clr3,   forget plot]
  table[row sep=crcr]{%
35.8	5.09099980856768\\
35.8	5.09999999994745\\
};
\addplot [color=clr9,   forget plot]
  table[row sep=crcr]{%
41	9.87360492796814\\
41	11.5420794477543\\
};
\addplot [color=clr4,   forget plot]
  table[row sep=crcr]{%
43.4	7.70311546550445\\
43.4	8.39289282036106\\
};
\addplot [color=clr3,   forget plot]
  table[row sep=crcr]{%
45.8	5.13099993171446\\
45.8	5.18273177906224\\
};
\addplot [color=clr9,   forget plot]
  table[row sep=crcr]{%
1	1.99429284312688\\
1	4.21764973505859\\
};
\addplot [color=clr4,   forget plot]
  table[row sep=crcr]{%
3.4	1.84044668928073\\
3.4	2.34397568271287\\
};
\addplot [color=clr3,   forget plot]
  table[row sep=crcr]{%
5.8	1.73321969832487\\
5.8	2.24287243789984\\
};
\addplot [color=clr9,   forget plot]
  table[row sep=crcr]{%
11	1.98196312429338\\
11	4.13142080823758\\
};
\addplot [color=clr4,   forget plot]
  table[row sep=crcr]{%
13.4	1.84902641525436\\
13.4	2.33164702758504\\
};
\addplot [color=clr3,   forget plot]
  table[row sep=crcr]{%
15.8	1.75358832045126\\
15.8	2.22552654477501\\
};
\addplot [color=clr9,   forget plot]
  table[row sep=crcr]{%
21	6.36666246997993\\
21	6.72695566656492\\
};
\addplot [color=clr4,   forget plot]
  table[row sep=crcr]{%
23.4	6.63905045817839\\
23.4	7.15042246341955\\
};
\addplot [color=clr3,   forget plot]
  table[row sep=crcr]{%
25.8	5.02999996509737\\
25.8	5.04799994646394\\
};
\addplot [color=clr9,   forget plot]
  table[row sep=crcr]{%
31	6.38950097273376\\
31	7.25792200143517\\
};
\addplot [color=clr4,   forget plot]
  table[row sep=crcr]{%
33.4	6.69606053336447\\
33.4	7.19479995320052\\
};
\addplot [color=clr3,   forget plot]
  table[row sep=crcr]{%
35.8	5.03099999990587\\
35.8	5.06499999995142\\
};
\addplot [color=clr9,   forget plot]
  table[row sep=crcr]{%
41	6.43902989072735\\
41	7.73784875423082\\
};
\addplot [color=clr4,   forget plot]
  table[row sep=crcr]{%
43.4	6.75295333287137\\
43.4	7.24321260877813\\
};
\addplot [color=clr3,   forget plot]
  table[row sep=crcr]{%
45.8	5.06299996600301\\
45.8	5.09599999722541\\
};
\addplot [color=clr9, forget plot]
  table[row sep=crcr]{%
0.7	6.61642465437895\\
1.3	6.61642465437895\\
};
\addplot [color=clr4, forget plot]
  table[row sep=crcr]{%
3.1	3.48522957647521\\
3.7	3.48522957647521\\
};
\addplot [color=clr3, forget plot]
  table[row sep=crcr]{%
5.5	3.29909771967089\\
6.1	3.29909771967089\\
};
\addplot [color=clr9, forget plot]
  table[row sep=crcr]{%
10.7	6.61158701824564\\
11.3	6.61158701824564\\
};
\addplot [color=clr4, forget plot]
  table[row sep=crcr]{%
13.1	3.55036574457205\\
13.7	3.55036574457205\\
};
\addplot [color=clr3, forget plot]
  table[row sep=crcr]{%
15.5	3.37606640715711\\
16.1	3.37606640715711\\
};
\addplot [color=clr9, forget plot]
  table[row sep=crcr]{%
20.7	10.7416425125128\\
21.3	10.7416425125128\\
};
\addplot [color=clr4, forget plot]
  table[row sep=crcr]{%
23.1	8.29058053647724\\
23.7	8.29058053647724\\
};
\addplot [color=clr3, forget plot]
  table[row sep=crcr]{%
25.5	5.09999999999146\\
26.1	5.09999999999146\\
};
\addplot [color=clr9, forget plot]
  table[row sep=crcr]{%
30.7	11.4982329188214\\
31.3	11.4982329188214\\
};
\addplot [color=clr4, forget plot]
  table[row sep=crcr]{%
33.1	8.29823291882136\\
33.7	8.29823291882136\\
};
\addplot [color=clr3, forget plot]
  table[row sep=crcr]{%
35.5	5.09999999994745\\
36.1	5.09999999994745\\
};
\addplot [color=clr9, forget plot]
  table[row sep=crcr]{%
40.7	11.5420794477543\\
41.3	11.5420794477543\\
};
\addplot [color=clr4, forget plot]
  table[row sep=crcr]{%
43.1	8.39289282036106\\
43.7	8.39289282036106\\
};
\addplot [color=clr3, forget plot]
  table[row sep=crcr]{%
45.5	5.18273177906224\\
46.1	5.18273177906224\\
};
\addplot [color=clr9, forget plot]
  table[row sep=crcr]{%
0.7	1.99429284312688\\
1.3	1.99429284312688\\
};
\addplot [color=clr4, forget plot]
  table[row sep=crcr]{%
3.1	1.84044668928073\\
3.7	1.84044668928073\\
};
\addplot [color=clr3, forget plot]
  table[row sep=crcr]{%
5.5	1.73321969832487\\
6.1	1.73321969832487\\
};
\addplot [color=clr9, forget plot]
  table[row sep=crcr]{%
10.7	1.98196312429338\\
11.3	1.98196312429338\\
};
\addplot [color=clr4, forget plot]
  table[row sep=crcr]{%
13.1	1.84902641525436\\
13.7	1.84902641525436\\
};
\addplot [color=clr3, forget plot]
  table[row sep=crcr]{%
15.5	1.75358832045126\\
16.1	1.75358832045126\\
};
\addplot [color=clr9, forget plot]
  table[row sep=crcr]{%
20.7	6.36666246997993\\
21.3	6.36666246997993\\
};
\addplot [color=clr4, forget plot]
  table[row sep=crcr]{%
23.1	6.63905045817839\\
23.7	6.63905045817839\\
};
\addplot [color=clr3, forget plot]
  table[row sep=crcr]{%
25.5	5.02999996509737\\
26.1	5.02999996509737\\
};
\addplot [color=clr9, forget plot]
  table[row sep=crcr]{%
30.7	6.38950097273376\\
31.3	6.38950097273376\\
};
\addplot [color=clr4, forget plot]
  table[row sep=crcr]{%
33.1	6.69606053336447\\
33.7	6.69606053336447\\
};
\addplot [color=clr3, forget plot]
  table[row sep=crcr]{%
35.5	5.03099999990587\\
36.1	5.03099999990587\\
};
\addplot [color=clr9, forget plot]
  table[row sep=crcr]{%
40.7	6.43902989072735\\
41.3	6.43902989072735\\
};
\addplot [color=clr4, forget plot]
  table[row sep=crcr]{%
43.1	6.75295333287137\\
43.7	6.75295333287137\\
};
\addplot [color=clr3, forget plot]
  table[row sep=crcr]{%
45.5	5.06299996600301\\
46.1	5.06299996600301\\
};
\addplot [color=clr9, forget plot]
  table[row sep=crcr]{%
0.4	4.21764973505859\\
0.4	6.12968998427154\\
1.6	6.12968998427154\\
1.6	4.21764973505859\\
0.4	4.21764973505859\\
};
\addplot [color=clr4, forget plot]
  table[row sep=crcr]{%
2.8	2.34397568271287\\
2.8	2.80943820974808\\
4	2.80943820974808\\
4	2.34397568271287\\
2.8	2.34397568271287\\
};
\addplot [color=clr3, forget plot]
  table[row sep=crcr]{%
5.2	2.24287243789984\\
5.2	2.66776191810686\\
6.4	2.66776191810686\\
6.4	2.24287243789984\\
5.2	2.24287243789984\\
};
\addplot [color=clr9, forget plot]
  table[row sep=crcr]{%
10.4	4.13142080823758\\
10.4	6.04127156228546\\
11.6	6.04127156228546\\
11.6	4.13142080823758\\
10.4	4.13142080823758\\
};
\addplot [color=clr4, forget plot]
  table[row sep=crcr]{%
12.8	2.33164702758504\\
12.8	2.82053930391937\\
14	2.82053930391937\\
14	2.33164702758504\\
12.8	2.33164702758504\\
};
\addplot [color=clr3, forget plot]
  table[row sep=crcr]{%
15.2	2.22552654477501\\
15.2	2.68861593309725\\
16.4	2.68861593309725\\
16.4	2.22552654477501\\
15.2	2.22552654477501\\
};
\addplot [color=clr9, forget plot]
  table[row sep=crcr]{%
20.4	6.72695566656492\\
20.4	8.35143424003738\\
21.6	8.35143424003738\\
21.6	6.72695566656492\\
20.4	6.72695566656492\\
};
\addplot [color=clr4, forget plot]
  table[row sep=crcr]{%
22.8	7.15042246341955\\
22.8	7.60819168686114\\
24	7.60819168686114\\
24	7.15042246341955\\
22.8	7.15042246341955\\
};
\addplot [color=clr3, forget plot]
  table[row sep=crcr]{%
25.2	5.04799994646394\\
25.2	5.08099999870758\\
26.4	5.08099999870758\\
26.4	5.04799994646394\\
25.2	5.04799994646394\\
};
\addplot [color=clr9, forget plot]
  table[row sep=crcr]{%
30.4	7.25792200143517\\
30.4	9.33318724276664\\
31.6	9.33318724276664\\
31.6	7.25792200143517\\
30.4	7.25792200143517\\
};
\addplot [color=clr4, forget plot]
  table[row sep=crcr]{%
32.8	7.19479995320052\\
32.8	7.63775210788479\\
34	7.63775210788479\\
34	7.19479995320052\\
32.8	7.19479995320052\\
};
\addplot [color=clr3, forget plot]
  table[row sep=crcr]{%
35.2	5.06499999995142\\
35.2	5.09099980856768\\
36.4	5.09099980856768\\
36.4	5.06499999995142\\
35.2	5.06499999995142\\
};
\addplot [color=clr9, forget plot]
  table[row sep=crcr]{%
40.4	7.73784875423082\\
40.4	9.87360492796814\\
41.6	9.87360492796814\\
41.6	7.73784875423082\\
40.4	7.73784875423082\\
};
\addplot [color=clr4, forget plot]
  table[row sep=crcr]{%
42.8	7.24321260877813\\
42.8	7.70311546550445\\
44	7.70311546550445\\
44	7.24321260877813\\
42.8	7.24321260877813\\
};
\addplot [color=clr3, forget plot]
  table[row sep=crcr]{%
45.2	5.09599999722541\\
45.2	5.13099993171446\\
46.4	5.13099993171446\\
46.4	5.09599999722541\\
45.2	5.09599999722541\\
};
\addplot [color=clr9, forget plot]
  table[row sep=crcr]{%
0.4	5.19097505186556\\
1.6	5.19097505186556\\
};
\addplot [color=clr4, forget plot]
  table[row sep=crcr]{%
2.8	2.54472720843624\\
4	2.54472720843624\\
};
\addplot [color=clr3, forget plot]
  table[row sep=crcr]{%
5.2	2.43055924251765\\
6.4	2.43055924251765\\
};
\addplot [color=clr9, forget plot]
  table[row sep=crcr]{%
10.4	5.1852631232602\\
11.6	5.1852631232602\\
};
\addplot [color=clr4, forget plot]
  table[row sep=crcr]{%
12.8	2.55502338999729\\
14	2.55502338999729\\
};
\addplot [color=clr3, forget plot]
  table[row sep=crcr]{%
15.2	2.43044467060721\\
16.4	2.43044467060721\\
};
\addplot [color=clr9, forget plot]
  table[row sep=crcr]{%
20.4	7.21441061771288\\
21.6	7.21441061771288\\
};
\addplot [color=clr4, forget plot]
  table[row sep=crcr]{%
22.8	7.37000560515888\\
24	7.37000560515888\\
};
\addplot [color=clr3, forget plot]
  table[row sep=crcr]{%
25.2	5.06399999785869\\
26.4	5.06399999785869\\
};
\addplot [color=clr9, forget plot]
  table[row sep=crcr]{%
30.4	8.1398092948876\\
31.6	8.1398092948876\\
};
\addplot [color=clr4, forget plot]
  table[row sep=crcr]{%
32.8	7.40873013323581\\
34	7.40873013323581\\
};
\addplot [color=clr3, forget plot]
  table[row sep=crcr]{%
35.2	5.07999999961264\\
36.4	5.07999999961264\\
};
\addplot [color=clr9]
  table[row sep=crcr]{%
40.4	8.60759218136711\\
41.6	8.60759218136711\\
};
\addlegendentry{Eq}

\addplot [color=clr4]
  table[row sep=crcr]{%
42.8	7.44855438555532\\
44	7.44855438555532\\
};
\addlegendentry{Rb}

\addplot [color=clr3]
  table[row sep=crcr]{%
45.2	5.11099999996318\\
46.4	5.11099999996318\\
};
\addlegendentry{Opt}

\addplot [color=black, draw=none, mark=+, mark options={solid, red}, forget plot]
  table[row sep=crcr]{%
nan	nan\\
};
\addplot [color=black, draw=none, mark=+, mark options={solid, red}, forget plot]
  table[row sep=crcr]{%
3.4	3.54895765743138\\
3.4	3.54996968110261\\
3.4	3.5526226251391\\
3.4	3.58995276629749\\
3.4	3.63962202426912\\
3.4	3.65778817588665\\
3.4	3.67729435816238\\
3.4	3.75861995472907\\
3.4	3.8825236385666\\
3.4	3.93044796843829\\
3.4	3.98727730234604\\
3.4	4.08834189924105\\
3.4	4.44115232569498\\
};
\addplot [color=black, draw=none, mark=+, mark options={solid, red}, forget plot]
  table[row sep=crcr]{%
5.8	3.32088273826803\\
5.8	3.32116976046307\\
5.8	3.33386430423859\\
5.8	3.34050180956638\\
5.8	3.35366983858268\\
5.8	3.36779589764962\\
5.8	3.38683737771147\\
5.8	3.39366726843456\\
5.8	3.41169067774089\\
5.8	3.41746698459143\\
5.8	3.50329121795577\\
5.8	3.50615584624974\\
5.8	3.53912424615981\\
5.8	3.55453721697122\\
5.8	3.62801438348252\\
5.8	3.64339924384175\\
5.8	3.79299140782628\\
5.8	3.90593926190617\\
5.8	4.15758119611913\\
};
\addplot [color=black, draw=none, mark=+, mark options={solid, red}, forget plot]
  table[row sep=crcr]{%
nan	nan\\
};
\addplot [color=black, draw=none, mark=+, mark options={solid, red}, forget plot]
  table[row sep=crcr]{%
13.4	3.56434156176438\\
13.4	3.60230192929179\\
13.4	3.6307453744945\\
13.4	3.66544095813679\\
13.4	3.67455425607575\\
13.4	3.68104630291905\\
13.4	3.71867252617718\\
13.4	3.74268864621699\\
13.4	3.98884429647746\\
13.4	4.04810284771731\\
};
\addplot [color=black, draw=none, mark=+, mark options={solid, red}, forget plot]
  table[row sep=crcr]{%
15.8	3.39640963642881\\
15.8	3.41343709082375\\
15.8	3.41616737126105\\
15.8	3.43510146179656\\
15.8	3.43768558134002\\
15.8	3.44220237538415\\
15.8	3.44383121996514\\
15.8	3.45047426242943\\
15.8	3.45624113962379\\
15.8	3.48085411472072\\
15.8	3.50977089098672\\
15.8	3.54146367844416\\
15.8	3.60343815653836\\
15.8	3.85156039876105\\
15.8	3.91307177337041\\
};
\addplot [color=black, draw=none, mark=+, mark options={solid, red}, forget plot]
  table[row sep=crcr]{%
21	11.030345069007\\
21	11.0306060014207\\
21	11.0315688890663\\
21	11.0352389500045\\
21	11.0407124813017\\
21	11.041357990674\\
21	11.0447956627356\\
21	11.0471999597017\\
21	11.0485533431128\\
21	11.0576227014919\\
21	11.0596842128157\\
21	11.0648551163284\\
21	11.0675657825114\\
21	11.0703205181595\\
21	11.0845608961085\\
21	11.0892134112547\\
21	11.0894695390783\\
21	11.4308687378615\\
21	11.4369555493202\\
21	11.4540801053665\\
21	11.4542819127632\\
21	11.455754594159\\
21	11.4629709927529\\
21	11.4637864903501\\
21	11.4718657525016\\
21	11.4817855212152\\
21	11.4905805364772\\
21	11.4970151470784\\
};
\addplot [color=black, draw=none, mark=+, mark options={solid, red}, forget plot]
  table[row sep=crcr]{%
23.4	8.32826520518113\\
23.4	8.34772415452764\\
23.4	8.35623072484061\\
23.4	8.37357514311285\\
23.4	8.40069995199641\\
23.4	8.43270772228411\\
23.4	8.44964625398629\\
23.4	8.52707737814519\\
23.4	8.53156888906627\\
23.4	8.54282873229114\\
23.4	8.6207602468538\\
23.4	8.69837282909289\\
23.4	8.73941264854944\\
23.4	8.81605787519364\\
23.4	9.20376166593351\\
};
\addplot [color=black, draw=none, mark=+, mark options={solid, red}, forget plot]
  table[row sep=crcr]{%
nan	nan\\
};
\addplot [color=black, draw=none, mark=+, mark options={solid, red}, forget plot]
  table[row sep=crcr]{%
nan	nan\\
};
\addplot [color=black, draw=none, mark=+, mark options={solid, red}, forget plot]
  table[row sep=crcr]{%
33.4	8.32478962624569\\
33.4	8.33524959177932\\
33.4	8.37850502278991\\
33.4	8.38551583125454\\
33.4	8.39347164492236\\
33.4	8.4096990405056\\
33.4	8.41759802723655\\
33.4	8.4256657955905\\
33.4	8.46346071492123\\
33.4	8.48477316288448\\
33.4	8.48893797710594\\
33.4	8.5432669954046\\
33.4	8.55518441899249\\
33.4	8.5617481229169\\
33.4	8.59591048479966\\
33.4	8.6220912428179\\
33.4	8.68168935160879\\
33.4	8.70159296825263\\
33.4	8.80848465966876\\
33.4	8.81118757576958\\
33.4	9.08333559149683\\
};
\addplot [color=black, draw=none, mark=+, mark options={solid, red}, forget plot]
  table[row sep=crcr]{%
35.8	5.13207225339372\\
35.8	5.20135122537872\\
35.8	5.20465731451361\\
35.8	5.32041843715814\\
35.8	5.38463367658062\\
};
\addplot [color=black, draw=none, mark=+, mark options={solid, red}, forget plot]
  table[row sep=crcr]{%
nan	nan\\
};
\addplot [color=black, draw=none, mark=+, mark options={solid, red}, forget plot]
  table[row sep=crcr]{%
43.4	8.40718651394879\\
43.4	8.41665627453269\\
43.4	8.42456060739902\\
43.4	8.43332791121108\\
43.4	8.44687921449355\\
43.4	8.44712306601807\\
43.4	8.47927237851398\\
43.4	8.57625336599628\\
43.4	8.58366090652539\\
43.4	8.60188201915981\\
43.4	8.72154320880053\\
43.4	8.73911791733576\\
43.4	8.74471448872466\\
43.4	8.77960850346686\\
43.4	8.84110125185587\\
43.4	9.33016456267796\\
43.4	9.35954609062869\\
};
\addplot [color=black, draw=none, mark=+, mark options={solid, red}, forget plot]
  table[row sep=crcr]{%
45.8	5.18796629087612\\
45.8	5.18803831961497\\
45.8	5.20038781378068\\
45.8	5.20058544432999\\
45.8	5.20494234715829\\
45.8	5.21054893705394\\
45.8	5.21603453964713\\
45.8	5.21607925221941\\
45.8	5.21651606793314\\
45.8	5.22938855265127\\
45.8	5.23009380092405\\
45.8	5.23796409102268\\
45.8	5.2448753210261\\
45.8	5.24895076579327\\
45.8	5.25122992093314\\
45.8	5.25659760165123\\
45.8	5.2566312085319\\
45.8	5.26396467992369\\
45.8	5.26498944878281\\
45.8	5.26948651882358\\
45.8	5.27328982629425\\
45.8	5.27481102033216\\
45.8	5.27792168103056\\
45.8	5.31050738921055\\
45.8	5.32751771945274\\
45.8	5.32802809860985\\
45.8	5.32947347536497\\
45.8	5.33132739636317\\
45.8	5.344206821906\\
45.8	5.34923505461326\\
45.8	5.35557573096891\\
45.8	5.36845157683856\\
45.8	5.37046745926011\\
45.8	5.37351260056887\\
45.8	5.4006779515581\\
45.8	5.41454982391117\\
45.8	5.42260514764299\\
45.8	5.43627356076313\\
45.8	5.44124265022357\\
45.8	5.4684211616335\\
45.8	5.46883291099677\\
45.8	5.4730300967384\\
45.8	5.5114776726996\\
45.8	5.51982821949996\\
45.8	5.52919142895399\\
45.8	5.5440489727627\\
45.8	5.55252029457733\\
45.8	5.56648427599125\\
45.8	5.57510155559525\\
45.8	5.5806334641306\\
45.8	5.59670714918972\\
45.8	5.70810993652195\\
45.8	5.71250839471745\\
45.8	5.79385108004576\\
45.8	5.80187972672386\\
45.8	5.82446288649875\\
45.8	5.84180668036391\\
45.8	5.91415794882711\\
45.8	5.97745121315593\\
45.8	6.16912091807241\\
45.8	6.17695115498428\\
45.8	6.31170624481677\\
45.8	6.40440411713728\\
};
\end{axis}
\end{tikzpicture}%

%% file: Conclusion.tex
\section{Conclusion}
\label{conclu}
We have studied the required time to download the data stored in a fog/cloud architecture, where multiple nodes providing different capabilities are available. We have modeled the corresponding multi-tier architecture by considering different performance parameters. On the one hand, we have introduced the download rate, $Rb$, and the communication delay, $d_{link}$, using values that are typical for 4G deployments. On the other hand, we have modeled the service delay of each node, i.e. the time required to process the corresponding data download request, $d_{service}$, with a $M/M/1$ queuing system. Furthermore, we have exploited NC as an enabler for the distributed data storage operation.

The corresponding system model boils down to a linear optimization problem, whose solution yields the optimal amount of data ($\alpha$) to store in (retrieve from) each of the nodes, taking into account the multiple parameters that characterize the system. Thus, we take full advantage of the properties of the whole system to yield the minimum download time. We have applied the proposed scheme in different scenarios and distribution methods to understand how the performance is affected by the system parameters. We have compared the behavior of the proposed scheme with that exhibited by other alternative solutions.

The use of multi-tier distributed data storage system clearly outperforms traditional single server solutions. On the other hand, when the distribution strategy considers system parameters, the download time is much lower than when an equal (fair) distribution is applied. We have also seen that the proposed scheme is able to keep a reasonable behavior even when some nodes suffer from rather poor conditions (either connectivity or overload). Under such circumstances, the results have shown that previously proposed solutions not only yield longer download times, but they also show more unpredictable performances.


In our future work we will broaden this work, by considering a dynamic architecture, where both system parameters and the number of available nodes might change. The estimation of the required storage data shall be carried out continuously, according to the system characteristics at each time instant. In addition, we plan to use more advanced optimization techniques, such as Model Predictive Control (MPC). In this sense, it is worth noting that the proposed optimization model could be easily extended to introduce different system parameters or additional constraints, for instance nodes having finite and heterogeneous storage capacities.

%% file: Appendix.tex
\appendix
\section{Optimization Problem}
\label{appendix}





Below we describe the solution of the optimization problem presented in Section \ref{min_problem}. Considering its convexity, the optimization problem can be reformulated in order to obtain a Linear Program (LP) \cite{Luenberger08} with constraints, by defining $z$ as the cost function. This reformulation maintains the previous properties, i.e. it is a convex and therefore, it has a global minimum. It can be represented as follows: 

\begin{equation}
\begin{aligned}
T^{*}_{total} =  &\min
& & z \\
& s.t.
& & z \geq d^{\text{request}}_i + \frac{\alpha_i k}{Rb_i}\\
&
& & \sum_{i=1}^{N}\alpha_{i}=1\\
&
& & 0 \preceq \boldsymbol{\alpha} \preceq 1
\end{aligned}
\label{eq:LP}
\end{equation}


\begin{remark}
We assume by physical coherence that all coefficients are strictly positive.
\end{remark}

The method of Lagrange Multipliers \cite{Rockafellar93} is used to find the solution for optimization problems constrained to one or more equalities. However, our LP also have inequalities, and thus, we need to extend the method to the Karush-Kuhn-Tacker (KKT) conditions \cite{Avriel03}. We can represent the LP obtained in Eq.~\ref{eq:LP} in a matrix form:

\begin{equation}
\begin{aligned}
& \min
& & z \\
& s.t.
& & A\alpha + \textbf{b}\leq \textbf{1}z\\
&
& & \sum_{i=1}^{N}\alpha_{i}=1\\
&
& & 0 \preceq \boldsymbol{\alpha} \preceq 1
\end{aligned}
\end{equation}

\noindent where $z\in \mathbb{R}$, $A\in\mathbb{R}^{N\times N}$, $\mathbf{b}\in \mathbb{R}^{N}$, $\alpha \in \mathbb{R}^{N}$. 

The Lagrangian function of our problem can be formulated as follows:

\begin{equation}
    L(z,\alpha,\lambda,\nu) = f(z,\alpha)+\sum_{j=1}^{m}\lambda_j g_j(z,\alpha) + \sum_{n=1}^{p}\nu_n h_n(z,\alpha)
\end{equation}

\noindent where $g_j$ are the functions corresponding to the inequality constraints, $h_n$ are the equality  functions for $j = 1,\dots,m$ y $n = 1,\dots,p$, and the parameters $\lambda$ and $\nu$ are the corresponding multipliers.

The derivative function of the Lagrangian is defined as: 
\begin{equation}
\begin{split}
   g(\lambda,\nu) = \underset{z,\alpha}{\text{inf}}\ L(z,\alpha,\lambda,\nu) = \nabla f(z,\alpha) + \sum_{j=1}^{m}\lambda_j \nabla g_j(z,\alpha) \\+ \sum_{n=1}^{l}\nu_n \nabla h_n(z,\alpha)
\end{split}
\end{equation}


Thus, the optimization problem can be defined as follows:

\begin{equation}
\begin{aligned}
& \text{max}
& & g(\lambda,\nu)\\
& \text{s.t.}
& &  \lambda \succeq 0
\end{aligned}
\end{equation}

Since there is a strong duality (Slater) \cite{Slater14} it can be established that the values that meet $g=0$ and $\lambda \succeq 0$ is the optimal solution to the optimization problem.

The optimization function for $\alpha \in \mathbb{R}^{N}$ and $i=1 \ldots N$ is the following:

\begin{equation}
\begin{aligned}
& \text{max}
& & z\\
& \text{s.t.}
& &  a_i\alpha_i+b_i-z\leq 0\\
&
& & \sum_{i=1}^{N}\alpha_i-1=0\\
&
& & -\alpha_i \leq 0\\
&
& & \alpha_i-1 \leq 0\\
\end{aligned}
\end{equation}

The Lagrangian of this optimization problem is formulated as:
\begin{equation}
\begin{split}
 L(\alpha,z,\lambda,\nu)=z+\sum_{i=1}^{N}\lambda_i(a_i\alpha_i+b_i-z) + \sum_{i=1}^{N}\lambda_{N+i}(-\alpha_i) + \\
 \sum_{i=1}^{N}\lambda_{2N+i}(\alpha_i-1) + \nu(\sum_{i=1}^{N}\alpha_i-1)
\end{split}
\end{equation}

Thus, it can be represented as the following equation system:

\begin{equation}
\begin{aligned}
& \frac{\partial L}{\partial z} = 1- \sum_{i=1}^{N}\lambda_i=0\\
& \frac{\partial L}{\partial \alpha_i} = a_i\lambda_i-\lambda_{N+i}+\lambda_{2N+i}+\nu=0\\
& \lambda_i (a_i\alpha_i+b_i-z)=0\\
& -\lambda_{N+i}\alpha_i = 0\\
& \lambda_{2N+i}(\alpha_i-1) = 0\\
& \nu (\sum_{i=1}^{N}\alpha_i-1)=0\\
\end{aligned}
\end{equation}